\documentclass[twoside,11pt]{article}

\usepackage[accepted]{melba}
\usepackage{mwe} % to get dummy images, only for the example
\usepackage{lscape}
\usepackage{amsmath,amsfonts}
\usepackage{multirow}
\usepackage{url}

\melbaheading{2022:026}{https://www.melba-journal.org/papers/2022:026.html}{2022}{1-54}{12/2021}{08/2022}{R. Mehta et al.}{}{}

\ShortHeadings{Benchmarking in QU-BraTS}{Mehta et al.}
\firstpageno{1}

\title{QU-BraTS: MICCAI BraTS 2020 Challenge on Quantifying Uncertainty in Brain Tumor Segmentation - Analysis of Ranking Scores and Benchmarking Results}

\author{{Raghav Mehta$^1$,
    Angelos Filos$^2$,
    Ujjwal Baid$^{3,4,5}$,
    Chiharu Sako$^{3,4}$,
    Richard McKinley$^6$,
    Michael Rebsamen$^6$,
    Katrin Dätwyler$^{6,53}$,
    Raphael Meier$^{54}$,
    Piotr Radojewski$^6$,
    Gowtham Krishnan Murugesan$^7$,
    Sahil Nalawade$^7$,
    Chandan Ganesh$^7$,
    Ben Wagner$^7$,
    Fang F. Yu$^7$,
    Baowei Fei$^8$,
    Ananth J. Madhuranthakam$^{7,9}$,
    Joseph A. Maldjian$^{7,9}$,
    Laura Daza$^{10}$,
    Catalina Gómez$^{10}$,
    Pablo Arbeláez$^{10}$,
    Chengliang Dai$^{11}$,
    Shuo Wang$^{11}$,
    Hadrien Reynaud$^{11}$,
    Yuanhan Mo$^{11}$,
    Elsa Angelini$^{12}$,
    Yike Guo$^{11}$,
    Wenjia Bai$^{11,13}$,
    Subhashis Banerjee$^{14,15,16}$,
    Linmin Pei$^{17}$,
    Murat AK$^{17}$,
    Sarahi Rosas-González$^{18}$,
    Ilyess Zemmoura$^{18,52}$,
    Clovis Tauber$^{18}$,
    Minh H. Vu$^{19}$,
    Tufve Nyholm$^{19}$,
    Tommy Löfstedt$^{20}$,
    Laura Mora Ballestar$^{21}$,
    Veronica Vilaplana$^{21}$,
    Hugh McHugh$^{22,23}$,
    Gonzalo Maso Talou$^{24}$,
    Alan Wang$^{22,24}$,
    Jay Patel$^{25,26}$,
    Ken Chang$^{25,26}$,
    Katharina Hoebel$^{25,26}$,
    Mishka Gidwani$^{25}$,
    Nishanth Arun$^{25}$,
    Sharut Gupta$^{25}$,
    Mehak Aggarwal$^{25}$,
    Praveer Singh$^{25}$,
    Elizabeth R. Gerstner$^{25}$,
    Jayashree Kalpathy-Cramer$^{25}$,
    Nicolas Boutry$^{27}$,
    Alexis Huard$^{27}$,
    Lasitha Vidyaratne$^{28}$,
    Md Monibor Rahman$^{28}$,
    Khan M. Iftekharuddin$^{28}$,
    Joseph Chazalon$^{29}$,
    Elodie Puybareau$^{29}$,
    Guillaume Tochon$^{29}$,
    Jun Ma$^{30}$,
    Mariano Cabezas$^{31}$,
    Xavier Llado$^{31}$,
    Arnau Oliver$^{31}$,
    Liliana Valencia$^{31}$,
    Sergi Valverde$^{31}$,
    Mehdi Amian$^{32}$,
    Mohammadreza Soltaninejad$^{33}$,
    Andriy Myronenko$^{34}$,
    Ali Hatamizadeh$^{34}$,
    Xue Feng$^{35}$,
    Quan Dou$^{35}$,
    Nicholas Tustison$^{36}$,
    Craig Meyer$^{35,36}$,
    Nisarg A. Shah$^{37}$,
    Sanjay Talbar$^{38}$,
    Marc-André Weber$^{39}$,
    Abhishek Mahajan$^{48}$,
    Andras Jakab$^{47}$,
    Roland Wiest$^{6,46}$
    Hassan M. Fathallah-Shaykh$^{45}$,
    Arash Nazeri$^{40}$,
    Mikhail Milchenko1$^{40,44}$,
    Daniel Marcus$^{40,44}$,
    Aikaterini Kotrotsou$^{43}$,
    Rivka Colen$^{43}$,
    John Freymann$^{41,42}$,
    Justin Kirby$^{41,42}$,
    Christos Davatzikos$^{3,4}$,
    Bjoern Menze$^{49,50}$,
    Spyridon Bakas$^*$$^{3,4,5}$,
    Yarin Gal$^*$$^2$,
    Tal Arbel$^*$$^{1,51}$}
    \\\\\\
    \scriptsize{
        $^{1}$Centre for Intelligent Machines (CIM), McGill University, Montreal, QC, Canada,
        $^{2}$Oxford Applied and Theoretical Machine Learning (OATML) Group, University of Oxford, Oxford, England,
        $^{3}$Center for Biomedical Image Computing and Analytics (CBICA), University of Pennsylvania, Philadelphia, PA, USA,
        $^{4}$Department of Radiology, Perelman School of Medicine at the University of Pennsylvania, Philadelphia, PA, USA,
        $^{5}$Department of Pathology and Laboratory Medicine, Perelman School of Medicine, University of Pennsylvania, Philadelphia, PA, USA,
        $^{6}$Support Center for Advanced Neuroimaging (SCAN), University Institute of Diagnostic and Interventional Neuroradiology, University of Bern, Inselspital, Bern University Hospital, Bern, Switzerland,
        $^{7}$Department of Radiology,  University of Texas Southwestern Medical Center, Dallas, TX, USA,
        $^{8}$Department of Bioengineering, University of Texas at Dallas, Texas, USA,
        $^{9}$Advanced Imaging Research Center, University of Texas Southwestern Medical Center, Dallas, TX, USA,
        $^{10}$Universidad de los Andes, Bogotá, Colombia,
        $^{11}$Data Science Institute, Imperial College London, London, UK,
        $^{12}$NIHR Imperial BRC, ITMAT Data Science Group, Imperial College London, London, UK,
        $^{13}$Department of Brain Sciences, Imperial College London, London, UK,
        $^{14}$Machine Intelligence Unit, Indian Statistical Institute, Kolkata, India,
        $^{15}$Department of CSE, University of Calcutta, Kolkata, India,
        $^{16}$ Division of Visual Information and Interaction (Vi2), Department of Information Technology, Uppsala University, Uppsala, Sweden,
        $^{17}$Department of Diagnostic Radiology, The University of Pittsburgh Medical Center, Pittsburgh, PA, USA,
        $^{18}$UMR U1253 iBrain, Université de Tours, Inserm, Tours, France,
        $^{19}$Department of Radiation Sciences, Umeå University, Umeå, Sweden,
        $^{20}$Department of Computing Science, Umeå University, Umeå, Sweden,
        $^{21}$Signal Theory and Communications Department, Universitat Politècnica de Catalunya, BarcelonaTech, Barcelona, Spain,
        $^{22}$Faculty of Medical and Health Sciences, University of Auckland, Auckland, New Zealand,
        $^{23}$Radiology Department, Auckland City Hospital, Auckland, New Zealand,
        $^{24}$Auckland Bioengineering Institute, University of Auckland, New Zealand,
        $^{25}$Athinoula A. Martinos Center for Biomedical Imaging, Department of Radiology, Massachusetts General Hospital, Boston, MA, USA,
        $^{26}$Massachusetts Institute of Technology, Cambridge, MA, USA,
        $^{27}$EPITA Research and Development Laboratory (LRDE), France,
        $^{28}$Vision Lab, Electrical and Computer Engineering, Old Dominion University, Norfolk, VA 23529, USA,
        $^{29}$EPITA Research and Development Laboratory (LRDE), Le Kremlin-Bicêtre, France,
        $^{30}$School of Science, Nanjing University of Science and Technology,
        $^{31}$Research Institute of Computer Vision and Robotics, University of Girona, Spain,
        $^{32}$Department of Electrical and Computer Engineering, University of Tehran, Iran,
        $^{33}$School of Computer Science, University of Nottingham, UK,
        $^{34}$NVIDIA, Santa Clara, CA, US,
        $^{35}$Biomedical Engineering, University of Virginia, Charlottesville, USA,
        $^{36}$Radiology and Medical Imaging, University of Virginia, Charlottesville, USA,
        $^{37}$Department of Electrical Engineering, Indian Institute of Technology - Jodhpur, Jodhpur, India,
        $^{38}$SGGS Institute of Engineering and Technology, Nanded, India,
        $^{39}$Institute of Diagnostic and Interventional Radiology, Pediatric Radiology and Neuroradiology, University Medical Center Rostock, Rostock, Germany
        $^{40}$Department of Radiology, Washington University, St. Louis, MO, USA,
        $^{41}$Leidos Biomedical Research, Inc, Frederick National Laboratory for Cancer Research, Frederick, MD, USA,
        $^{42}$Cancer Imaging Program, National Cancer Institute, National Institutes of Health, Bethesda, MD, USA,
        $^{43}$Department of Diagnostic Radiology, University of Texas MD Anderson Cancer Center, Houston, TX, USA,
        $^{44}$Neuroimaging Informatics and Analysis Center, Washington University, St. Louis, MO, USA,
        $^{45}$Department of Neurology, The University of Alabama at Birmingham, Birmingham, AL, USA,
        $^{46}$Institute for Surgical Technology and Biomechanics, University of Bern, Bern, Switzerland,
        $^{47}$Center for MR-Research, University Children’s Hospital Zurich, Zurich, Switzerland,
        $^{48}$Tata Memorial Centre, Homi Bhabha National Institute, Mumbai, India,
        $^{49}$Department of Quantitative Biomedicine, University of Zurich, Zurich, Switzerland,
        $^{50}$Department of Informatics, Technical University of Munich, Munich, Germany,
        $^{51}$MILA - Quebec Artificial Intelligence Institute, Montreal, QC, Canada,
        $^{52}$Neurosurgery department, CHRU de Tours, Tours, France,
        $^{53}$Human Performance Lab, Schulthess Clinic, Zurich, Switzerland,
        $^{54}$armasuisse S+T, Thun, Switzerland.
    }
    \\\\
    $^*$ Senior Authors
    \\\\
    Corresponding author: Raghav Mehta. Email - \url{raghav@cim.mcgill.ca}
}

\begin{document}

% top matter
\maketitle

% abstract
\begin{abstract}%   
Deep learning (DL) models have provided state-of-the-art performance in various medical imaging benchmarking challenges, including the Brain Tumor Segmentation (BraTS) challenges. However, the task of focal pathology multi-compartment segmentation (e.g., tumor and lesion sub-regions) is particularly challenging, and potential errors hinder translating DL models into clinical workflows. Quantifying the reliability of DL model predictions in the form of uncertainties could enable clinical review of the most uncertain regions, thereby building trust and paving the way toward clinical translation. Several uncertainty estimation methods have recently been introduced for DL medical image segmentation tasks. Developing scores to evaluate and compare the performance of uncertainty measures will assist the end-user in making more informed decisions. In this study, we explore and evaluate a score developed during the BraTS 2019 and BraTS 2020 task on uncertainty quantification (QU-BraTS) and designed to assess and rank uncertainty estimates for brain tumor multi-compartment segmentation. This score (1) rewards uncertainty estimates that produce high confidence in correct assertions and those that assign low confidence levels at incorrect assertions, and (2) penalizes uncertainty measures that lead to a higher percentage of under-confident correct assertions. We further benchmark the segmentation uncertainties generated by 14 independent participating teams of QU-BraTS 2020, all of which also participated in the main BraTS segmentation task. Overall, our findings confirm the importance and complementary value that uncertainty estimates provide to segmentation algorithms, highlighting the need for uncertainty quantification in medical image analyses. Finally, in favor of transparency and reproducibility, our evaluation code is made publicly available at~\url{https://github.com/RagMeh11/QU-BraTS}.
\end{abstract}

% keywords

% Introduction (or first section)
\section{Introduction}
Machine learning groups often struggle to gain access to large-scale annotated medical imaging datasets for training and testing their algorithms. As a result, many researchers rely on smaller proprietary datasets, making it challenging to show the full potential of their algorithms and even more challenging to compare their results against other published methods. Therefore, medical image analysis challenges \citep{BraTS2015, BraTS2018, simpson2019large, antonelli2021medical, kurc2020segmentation, refugechallenge, isicchallenge, acdcchallenge, isegchallenge, kitschallenge, pati2021federated, IMAGECLEF, VISCERAL, LIVER2007} play a pivotal role in developing machine learning algorithms for medical image analysis by making large-scale, carefully labeled, multi-center, real-world datasets publicly available for training, testing, and comparing machine learning algorithms. In particular, the Brain Tumor Segmentation (BraTS) challenge has provided the community with a benchmarking platform to compare segmentation methods for over ten years, increasing the dataset size each year \citep{BraTS2015, bakas2017advancing, BraTS2018, baid2021rsna}. The availability of the dataset, and the challenge itself, have permitted the development of many new successful deep learning based approaches such as the DeepMedic \citep{kamnitsas2016deepmedic, DeepMedic} and the nnU-Net \citep{nn-Unet, isensee2021nnuNatureMethods}. \\
 
Despite their success in many medical image analysis challenges, the resulting deep learning algorithms are typically not translated into the clinical setting for various reasons. One problem is that most deep learning models produce deterministic outputs. That is, they do not communicate the uncertainties associated with their predictions. This is problematic in the challenging context of segmenting pathological structures (e.g., tumors, lesions), as even the top-performing methods produce errors. Providing uncertainties associated with the machine learning predictions could permit the end-user (e.g., clinician) to review and correct the model predictions where the model is not certain about its predictions. \\ 

Bayesian Deep Learning provides a popular framework to allow deep learning models to generate predictions and their associated uncertainties ~\citep{BDLneal,UncReview}. Recent advances \citep{MCD, DeepEnsemble, VarUnc, WeightUnc} in Bayesian Deep Learning have led to widespread adaptations of the frameworks for different tasks in medical image analysis \citep{UltraUnc, UDA, VoxelMorph}. There are recent attempts to evaluate uncertainties associated with model predictions in medical image analysis \citep{BUNet, UDR, UncChest, UncMemo, UncDRBench, QUBIQ}, which quantifies whether these uncertainties can adequately capture model confidence in these domains. To the best of our knowledge, there has not yet been a single unified approach to evaluate the model uncertainties in the context of medical image segmentation. \\

The main focus of this work is three-fold: i) to develop an uncertainty evaluation score with a down-stream clinical goal in mind; ii) to benchmark the various participating teams from a recent BraTS challenge \citep{BraTS2018}, using the developed evaluation score; and iii) to make the associated evaluation code publicly available for future benchmarking of Bayesian Deep Learning methods for medical image segmentation. In particular, we focus on developing an uncertainty evaluation criterion for brain tumor segmentation. We aim to develop a Computer-Aided Diagnosis (CAD) system where the pathology size is smaller than the surrounding healthy tissue. In this context, the objectives are that the uncertainty estimates associated with an automatic segmentation system reflect that the system is (a) confident when correct and (b) uncertain when incorrect. These criteria would mainly permit uncertain predictions to be flagged and brought to the attention of the clinical expert, rather than overburdening the expert by having to review the entirety of the prediction. To this end, we present the resulting uncertainty evaluation score \citep{midl-metric} and the rankings and results for 14 teams participating in the Quantification of Uncertainty for Brain Tumor Segmentation (QU-BraTS) 2020 challenge. The various analyses of the methods and results produced by the different teams highlight the necessity of the different components of our developed score. The results indicate that the segmentation results and the associated uncertainties give complementary information as teams performing well on one task do not necessarily perform well on the other. Qualitative results show that the developed score measures the desired real-world properties for tumor segmentation uncertainties.  \\

\section{Related Work}
\label{sec:related}

Recent works~\citep{WhatUnc, AleaOrEpis} show that uncertainties associated with the outputs of a machine learning model are primarily divided into two sub-types: (i) Epistemic uncertainty, which captures the uncertainty associated with the model parameters, and (ii) Aleatoric uncertainty, which captures the uncertainty inherent in the data. The epistemic uncertainty captures our ignorance about which model generated our collected data. This uncertainty can be reduced to zero if the model is provided with an infinite amount of data, permitting the model parameters to learn the true distribution of the data generation model. The aleatoric uncertainty could result from measurement noise, for example, and therefore cannot be reduced even with more data collection. Both epistemic and aleatoric uncertainties play essential roles in medical image analysis. Epistemic uncertainty indicates where to trust the model output~\citep{BUNet, UDR}, and aleatoric uncertainty reflects the prevalent noise in the data \citep{USQE}.\\

Several recent papers~\citep{UncChest, BUNet, UncMemo, BSN} show cases where uncertainty estimates correlate with errors in a machine learning model. These results show promise that estimating uncertainties make a better adaptation of deep learning models in real-world scenarios possible. However, in the medical image analysis field, to date, there is an unmet need to (1) systemically quantify and compare how well different uncertainty estimates properly communicate the degree of confidence in the output and (2) to rank the performance of competing estimates, given the objectives of the task and the requirements during a clinical review. \\

The most popular metrics for measuring model confidence output are the expected calibration error (ECE) and the maximum calibration error (MCE)~\citep{BatchEnsemble, pitfalls}. These metrics are useful for quantitatively measuring model calibration. However, these metrics are based on the softmax probabilities. Furthermore, a simple post-processing technique like temperature scaling \citep{TS} can make a deterministic and a probabilistic model equally calibrated. ECE and MCE metrics cannot differentiate between these temperature-calibrated models. \\

In another paper, \cite{DeepEnsemble} evaluate the usefulness of the predictive uncertainty for decision making by evaluating the model output only in cases where the model’s confidence is above a user-specified threshold. Their main idea is that if the confidence estimates are well-calibrated on the data distribution seen during training, one can trust the model predictions when the reported confidence is high and result to a different solution when the model is not confident. They showed that when the model is evaluated on its most confident prediction, the model accuracy is high compared to when the model is evaluated on all outputs. Though this is encouraging and allows for the comparison of different uncertainty generation methods, it does not consider how many model outputs were discarded at a certain threshold. Using this criterion, a model which has low accuracy but high uncertainty when evaluated on all predictions is rewarded. This model is undesirable in a practical scenario, and leading to the rejection of most of its predictions to achieve high accuracy.\\

\cite{ESS} designed a metric to quantify uncertainty for the task of semantic segmentation. They made the following assumption during the metric design: if a model is confident about its prediction, it should be accurate, which implies that if a model is inaccurate on output, it should be uncertain. With this in mind, they calculate the following two probabilities at different uncertainty thresholds: (i) p(accurate$|$certain): the probability that the model is accurate on its output given that it is confident; (ii) p(uncertain$|$inaccurate): the probability that the model is uncertain about its output given that it has made a mistake in its prediction (i.e., is inaccurate). They used the metric to compare different BDL methods for the semantic segmentation task. Though this metric is useful for semantic segmentation, where each pixel in an image is labelled as one class, it is not useful for the task of pathology segmentation where there is a high class-imbalance problem and the number of pixels (voxels) of interest (pathology) is low compared to the background-healthy class. For example, in the brain tumour segmentation task, 99.9\% of the voxels belong to the background (healthy tissue) while only 0.1\% belongs to the foreground (pathology). Due to a high class imbalance, p(accurate$|$certain) would be dominated by healthy (background) voxels, most of which can be accurately classified with high certainty. \\

\cite{PQD} developed a metric, Probability-based Detection Quality (PDQ), to evaluate the uncertainty estimate for the object detection task. The authors combine the class labelling measure (i.e., label quality) and the bounding box detection measure (i.e., spatial quality) into the metric. Here, spatial quality measures how well the detection describes where the object is within the image. Label quality measures how effectively a detection identifies the object class. These are averaged over all possible combinations of bounding boxes and labels generated using multiple samples. The authors also organized a \href{https://nikosuenderhauf.github.io/roboticvisionchallenges/object-detection.html}{challenge} associated with this task at the Annual Conference on Computer Vision and Pattern Recognition (CVPR) 2019. The paper and its associated challenge~\citep{PODC} illustrate the importance of developing uncertainty quantification metrics that are tailored to the task of interest.\\
 
\cite{UBTS} made the first step towards quantifying uncertainty for the brain tumor segmentation task. They compared various uncertainty generation methods such as MC-Dropout, Deep Ensemble \citep{MCD, DeepEnsemble}, and others, using the standard metrics like ECE, MCE, and reliability diagrams. In addition, they proposed a new metric, Uncertainty-Error (U-E) overlap. The results showed that Deep Ensemble could produce more reliable uncertainty measures than other methods.\\

\section{Uncertainty Evaluation Score}
\label{UncMetric}
This work focuses on the publicly available BraTS challenge dataset~\citep{BraTS2015, BraTS2018, bakas2017advancing}, which consists of both High-Grade Glioma (HGG) and Low-Grade Glioma (LGG) cases, as described in the previous BraTS manuscripts and on The Cancer Imaging Archive (TCIA) \citep{TCIA}. However, this naming convention is now obsolete. Following the 2021 World Health Organization (WHO) classification of tumors of the central nervous system (CNS) \citep{LouisWho2021}, the data provided by the BraTS challenge should be described as including: 1) adult-type diffuse gliomas, 2) pediatric-type diffuse low-grade gliomas, and 3) pediatric-type high-grade gliomas. The adult-type diffuse gliomas in the BraTS dataset comprise Glioblastoma (IDH-wildtype, CNS WHO grade 4) and Astrocytoma (IDH-mutant, CNS WHO grades 2-4). Ground truth labels generated and signed off by clinical experts are provided for each patient case and consist of 3 tumor sub-regions: enhancing tumor core, necrotic core, and peritumoral edematous/infiltrated tissue (here onward referred to as edema) \citep{BraTS2018}. However, focusing on the clinical relevance, the submitted algorithms are not evaluated on each of these tumor sub-regions but on higher-level tumor entities that relate directly to the surgical and radiotherapy importance. Specifically, the tumor entities considered during the evaluation and ranking of algorithms are: (i) the enhancing tumor core (ET), (ii) the tumor core (TC), which consists of the union of ET and the necrotic core, and (iii) the whole tumor (WT), which consists of all three sub-regions namely edema, necrotic core, and enhancing tumor core, and radiographically is defined by the abnormal FLAIR signal envelope. The performance of the automatic segmentation methods is finally evaluated using the Dice Similarity coefficient (referred to as $DSC$ from here onward) and the 95th percentile of the Hausdorff distance between the predicted labels and the provided ground truth. \\

The objective of the uncertainty quantification task was to evaluate and rank the uncertainty estimates for the task of brain tumor segmentation. To this end, each team provided their output labels for the multi-class segmentation task and the estimated voxel-wise uncertainties for each of the associated tumor entities, namely, WT, TC, and ET. These uncertainties were required to be normalized in the range of $0-100$ for ease of computation. For each tumor entity, the uncertain voxels were filtered at $N$ predetermined uncertainty threshold values $\tau_{1,..,N}$, and the model performance was assessed based on the metric of interest (i.e., the $DSC$ in this case) of the remaining voxels at each of these thresholds ($\tau_{1,..,N}$). For example, $\tau = 75$ implies that all voxels with uncertainty values $\ge 75$ are marked as uncertain, and the associated predictions are filtered out and not considered for the subsequent $DSC$ calculations. In other words, the $DSC$ values are calculated for the remaining predictions of the unfiltered voxels. This evaluation rewards models where the confidence in the incorrect assertions (i.e., False Positives, denoted FPs, and False Negatives, denoted FNs) is low and high for correct assertions (i.e., True Positives, denoted TPs, and True Negatives, denoted TNs). For these models, it is expected that as more uncertain voxels are filtered out, the $DSC$ score, calculated only on the remaining unfiltered voxels, increases. \\

Although the criterion mentioned above helps measure performance in terms of $DSC$, the metric of interest, it does not keep track of the total number of filtered voxels at each threshold. In real practice, an additional penalty should be provided to a system that filters out many voxels at a low threshold to achieve high performance on the metric of interest, as it will increase the reviewing burden on clinical raters. One solution is to add a penalty based on the total number of filtered voxels at each uncertainty threshold. This strategy is also not ideal as it will also penalize methods that filter out FPs/FNs, areas where mistakes are made. Instead, the evaluation method chosen penalizes methods that filter out only the correctly predicted voxels (i.e., TP and TN). Given that the specific tumor segmentation task has a high-class imbalance between pathological and healthy tissue, different penalties are assigned to TPs and TNs. The ratio of filtered TPs (FTP) is estimated at different thresholds ($\tau_{1,..,N}$) and is measured relative to the unfiltered values ($\tau = 100$) such that FTP = ($\text{TP}_{\text{100}}$ - $\text{TP}_\tau$) / $\text{TP}_{\text{100}}$. The ratio of filtered TNs is calculated similarly. This evaluation essentially penalizes approaches that filter out a large percentage of TP or TN relative to $\tau=100$ voxels (i.e., more uncertain about correct assertions) to attain the reported $DSC$ value, thereby rewarding approaches with a lower percentage of uncertain TPs/TNs.\\

\begin{figure*}[t]
\centering
\includegraphics[width=1.00\textwidth]{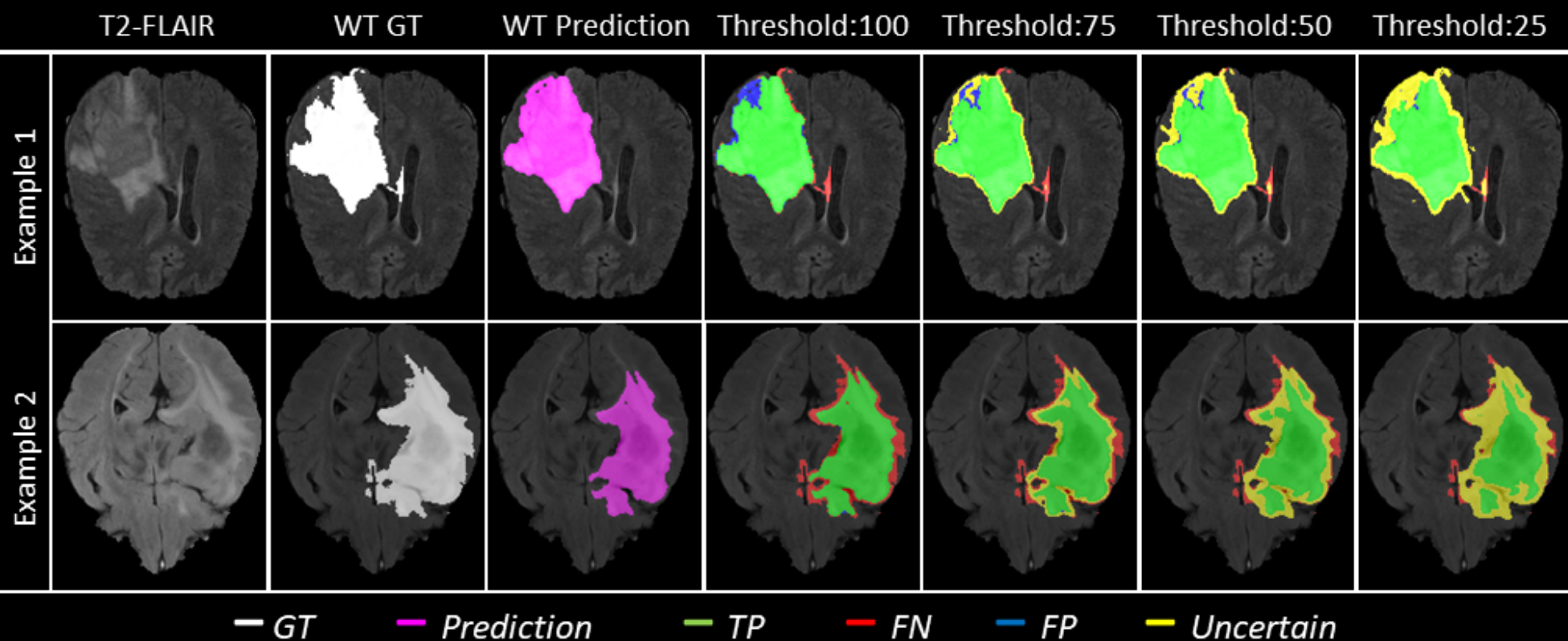}
\center \caption{Effect of uncertainty thresholding on two different example patient MRI slices (Row-1 and Row-2) for whole tumor (WT) segmentation. (a) T2-FLAIR MRI. (b) WT Ground Truth (c) Overall Model Prediction (d) Results with No filtering, Uncertainty Threshold = 100. (e) Uncertainty Threshold = 75 (f) Uncertainty Threshold = 50 (g) Uncertainty Threshold = 25. It is desired that with decrease in the uncertainty threshold, more False Positives (blue) and False Negative (red) voxels are filtered out (marked as uncertain - yellow) while True Positive (green) and True Negative voxels remain unfiltered.}
\label{fig:UncExam}
\end{figure*}

\begin{table}[t]
\centering
\caption{Change in $DSC$, Filtered True Positives (FTP) ratio, and Filtered True Negatives (FTN) ratio with change in uncertainty thresholds for two different example slices shown in Figure~\ref{fig:UncExam}.}

\begin{tabular}{|l|l|l|l|l|}
\hline
\multirow{2}{*}{}  & \multicolumn{4}{c|}{\textbf{$DSC$}}                                                            \\ \cline{2-5} 
                   & \textbf{$DSC$ at 100 (baseline)} & \textbf{$DSC$ at 75} & \textbf{$DSC$ at 50} & \textbf{$DSC$ at 25} \\ \hline
\textbf{Example-1} & 0.94                  & 0.96                  & 0.965                 & 0.97                  \\ \hline
\textbf{Example-2} & 0.92                  & 0.955                 & 0.97                  & 0.975                 \\ \hline
\multirow{2}{*}{}  & \multicolumn{4}{l|}{\textbf{Ratio of Filtered TPs (1 - ($\text{TP}_x$ / $\text{TP}_{\text{baseline ($\tau$=100)}}$))}}                \\ \cline{2-5} 
                   & \textbf{FTP at 100}  & \textbf{FTP at 75}  & \textbf{FTP at 50}  & \textbf{FTP at 25}  \\ \hline
\textbf{Example-1} & 0.00                  & 0.00                  & 0.05                  & 0.1                   \\ \hline
\textbf{Example-2} & 0.00                  & 0.00                  & 0.15                  & 0.25                  \\ \hline
\multirow{2}{*}{}  & \multicolumn{4}{l|}{\textbf{Ratio of Filtered TNs (1 - ($\text{TN}_x$ / $\text{TN}_{\text{baseline ($\tau$=100)}}$))}}                \\ \cline{2-5} 
                   & \textbf{FTN at 100}  & \textbf{FTN at 75}  & \textbf{FTN at 50}  & \textbf{FTN at 25}  \\ \hline
\textbf{Example-1} & 0.00                  & 0.0015                & 0.0016                & 0.0019                \\ \hline
\textbf{Example-2} & 0.00                  & 0.0015                & 0.0026                & 0.0096                \\ \hline
\end{tabular}%}
\label{tab:unc_exam}
\end{table}

Figure~\ref{fig:UncExam} and Table~\ref{tab:unc_exam} depict qualitative examples and their associated quantitative results. Here, decreasing the threshold  ($\tau$) leads to filtering out voxels with incorrect assertions. This filtering, in turn, leads to an increase in the $DSC$ value for the remaining voxels. Example 2 indicates a marginally better $DSC$ value than the slice in example 1 at uncertainty thresholds ($\tau$) 50 and 25. However, the Ratio of FTPs and FTNs indicates that this is at the expense of marking more TPs and TNs as uncertain. \\

To ensure that the generated output segmentations are directly associated with the BraTS challenge protocol, the generated uncertainties are expected to be produced for these "binary" tumor entities, i.e., ET, TC, and WT. The associated uncertainties are evaluated using the scores defined above for each tumor entity.\\

Finally, the resulting uncertainty measures for each team are ranked according to a unified score which combines the area under three curves: 1) $DSC$ vs $\tau$, 2) FTP vs $\tau$, and 3) FTN vs $\tau$, for different values of $\tau$. The unified score is calculated as follows: 

\begin{equation}
\text{score}_\text{tumor\_entity} = \frac{{AUC}_1 + (1-{AUC}_2) + (1-{AUC}_3)}{3}.
\label{Equation-score}
\end{equation}

In the context of the BraTS uncertainty evaluation task (QU-BraTS), the score is estimated for each tumor entity separately and then used to rank the participating methods.  

\subsection{A 3D U-Net Based Experiment} \label{3DUNetExp}

\begin{figure*}[t]
\centering
\includegraphics[width=\textwidth]{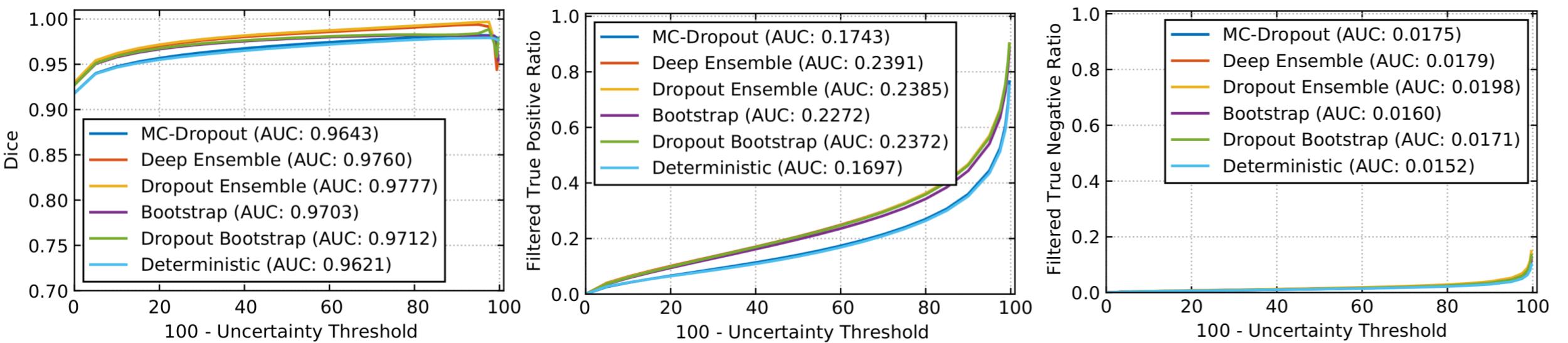} 
\center \caption{Effect of changing uncertainty threshold ($\tau$) on WT for entropy measure. Specifically, we plot (left) $DSC$, (middle) Filtered True Positive Ratio, and (right) Filtered True Negative Ratio as a function of 100 - $\tau$. We plot the curves for six different uncertainty generation methods, namely, MC-Dropout, Deep Ensemble, Dropout Ensemble, Bootstrap, Dropout Bootstrap, and Deterministic. All methods use entropy as a measure of uncertainty.}
\label{fig:UncEntropy}
\end{figure*}

Experiments were devised to show the functioning of the derived uncertainty evaluations and rankings. A modified 3D U-Net architecture \citep{3DUNet,mehta20183d} generates the segmentation outputs and corresponding uncertainties. The network was trained ($n=228$), validated ($n=57$), and tested ($n=50$) based on the publicly available BraTS 2019 training dataset ($n=335$) \citep{BraTS2015, BraTS2018, bakas2017advancing, tciaGbmSeg2017, tciaLggSeg2017}. The performances of WT segmentation with the entropy uncertainty measure \citep{UNAL}, which captures the average amount of information contained in the predictive distribution,  are shown in Figure~\ref{fig:UncEntropy}. Here uncertainties are estimated using MC-Dropout \citep{MCD}, Deep Ensemble \citep{DeepEnsemble}, Dropout Ensemble \citep{EMCD}, Bootstrap, Dropout Bootstrap, and a Deterministic softmax entropy measure. Dropout bootstrap shows the best $DSC$ performance (highest AUC) and has the worst performance for FTP and FTN curves (highest AUC). This result shows that the higher performance in $DSC$ is at the expense of a higher number of filtered correct voxels. Overall, the score is working in line with the objectives. However, there is no clear winner amongst these uncertainty methods in terms of rankings. \\

\section{BraTS 2020 Quantification of Uncertainty (QUBraTS) challenge -- Materials and Methods}

\subsection{Dataset}

The BraTS 2020 challenge dataset \citep{BraTS2015, BraTS2018, bakas2017advancing, tciaGbmSeg2017, tciaLggSeg2017} is divided into three cohorts: Training, Validation, and Testing. The Training dataset is composed of multi-parametric MRI (mpMRI) scans from 369 diffuse glioma patients. Each mpMRI set contains four different sequences: native T1-weighted (T1), post-contrast T1-weighted (T1ce), T2-weighted (T2), and T2 Fluid-Attenuated-Inversion-Recovery (FLAIR). Each MRI volume is skull-stripped (also known as brain extraction) \citep{thakur2020brain}, co-aligned to a standard anatomical atlas (i.e., SRI24 \citep{rohlfing2010sri24}), and resampled to $1mm^3$ voxel resolution. Expert human annotators provided GT tumor labels, consisting of 3 classes described previously. Note that there is no "ground-truth" uncertainty label.\\

The BraTS 2020 Validation cohort is composed of 125 cases of patients with diffuse gliomas. Similar to the training dataset, this also contains four different mpMRI sequences for each case. The validation dataset allows participants to obtain preliminary results in unseen data and their cross-validated results on the training data. The GT labels for the validation data are not provided to the participants. \\

The BraTS 2020 Testing cohort is then used for the final ranking of the participating team. It is comprised of a total of 166 cases. The exact type of glioma is not revealed to the participating teams. Each team gets a window of 48 hours to upload their results to the challenge evaluation platform (\url{https://ipp.cbica.upenn.edu/}) \citep{davatzikos2018cancer}.\\

\subsection{Evaluation framework}

The University of Pennsylvania Image Processing Portal (\url{https://ipp.cbica.upenn.edu/}) is used to evaluate all BraTS participating algorithms quantitatively. This portal allows the registration of new teams to access the BraTS datasets and the framework for automatically evaluating all participating algorithms on all three (i.e., training, validation, and testing) cohorts\footnote{Access to the BraTS testing datasets is not possible after the conclusion of the challenge.}. In addition to the IPP, and in favor of reproducibility and transparency, we implement the quantitative evaluation of uncertainty publicly available through GitHub\footnote{\url{https://github.com/RagMeh11/QU-BraTS}}. As mentioned previously, the evaluation framework expects the challenge participants to provide multi-class brain tumor segmentation labels and their associated voxel-wise uncertainties for three tumor entities: whole tumor (WT), tumor core (TC), and enhancing tumor (ET). These uncertainties are expected to be normalized between 0-100 for ease of computation.

\subsection{Participating Methods}
In total, 14 teams participated in the QU-BraTS 2020 challenge. All teams utilized a Convolutional Neural Network (CNN) based approach for the tumor segmentation task and the generation of associated uncertainty maps. Detailed descriptions of 12/14 proposed approaches are given below\footnote{Two teams, namely Frankenstein \citep{Frankenstein} and NSU-btr \citep{nsu-btr}, withdrew from participating in this paper.}. Details regarding the CNN segmentation architectures utilized by each team are not described in detail here, as this paper focuses on uncertainty generation methods rather than the segmentation itself. Readers are requested to refer to each team's individual papers (as cited below) for more details about the CNN architecture used for the segmentation task. A preliminary version of the QU-BraTS challenge was run in conjunction with the BraTS 2019 challenge. Appendix B provides details about the participating teams and their performance. We did not include the analysis results of the QU-BraTS 2019 challenge. The task was run as a preliminary task without employing any statistical significance analysis ranking scheme to evaluate the participating teams.

\subsubsection{Method-1: Team SCAN \citep{SCAN}} \label{TeamSCAN}
The method uses the DeepSCAN \citep{SCAN2020} model. The training of the model was performed using a combination of focal loss \citep{FocalLoss} and a Kullback-Leibler divergence: for each voxel and each tumor entity, the model produces an output $p \in (0,1)$ (corresponding to the output of a standard binary classification network) and an output $q \in (0,0.5)$ which represents the probability that the classifier output differs from the ground truth on that tumor entity. The probability $q$ is supervised by the label $z$, which is the indicator function for disagreement between the classifier (thresholded at the $p = 0.5$ level) and the ground truth. Given $q$, an annealed version of the ground truth is formed, $w = (1-x) \cdot q + x \cdot (1-q)$. Focal KL divergence between $w$ and $p$ is defined as follows: $$ \text{Focal}_\text{KL} (w||p) = (p-w)^2 (w \cdot log(w) - w \cdot log(p)).$$ The final loss function is given by: $$ \text{Loss} = 0.1 \cdot \text{Focal}(p,x) + 0.9 \cdot \text{Focal}_\text{KL}(w||p) + 0.9  \cdot  \text{BCE}(q, z).$$ An ensemble of the networks was utilized in the final output, where from different predictions, $p$ and $q$ were combined to a single probability $q  \cdot  I_{p\le0.5} + (1 - q)I_{p\ge0.5}$. The final uncertainty output (denoted $q$ above) was normalized into the range of 0 to 100: $100*(1-2q)$. The uncertainty in the ensemble can likewise be extracted as for any ordinary model with a sigmoid output $x$ as: $100  \cdot  (1 - 2|0.5 - x|)$ \\

While this uncertainty measure gives a measure of uncertainty both inside and outside the provided segmentation, it was empirically observed that treating all positive predictions as certain and only assigning uncertain values to only negative predictions gives better performance on the challenge scores.

\subsubsection{Method-2: Team Alpaca \citep{ALPACA}} \label{TeamAlpaca}
An ensemble of three different 2D segmentation networks \citep{DenseNET-169, SERESNEXT-101, SENet-154} was used. The softmax probabilities from each of the three networks were averaged to generate the final probability maps. These probability maps were used to generate the uncertainty maps for each tumor entity. This was computed by mapping the most confident prediction value to 0 and the least confident value to 100.

\subsubsection{Method-3: Team Uniandes \citep{UNIANDES}}
A novel deep learning architecture named Cerberus was proposed. The uncertainty maps were produced by taking the compliment of the final segmentation softmax probability maps, and rescaling them between 0 and 100.

\subsubsection{Method-4: Team DSI\_Med \citep{DSIMed}}
Five attention-gated U-Net models were trained. The uncertainty maps were normalised between 0 and 100 for the four nested tumor entities. For each uncertainty map, the maximum softmax probability from the five models for each voxel in each entity was taken. The voxels were either part of the given nested entity or not, judging by the segmentation maps acquired from the ensemble of five models. The probabilities of those voxels that belong to the nested entity were inverted and multiplied by 100. The results were then rounded to get into the 0-100 range.\\

Double thresholds were further applied to refine the uncertainty maps. Low and high probability thresholds for each nested entity were empirically defined: WT(0.1, 0.3), TC(0.2, 0.3) ET(0.3, 0.5). For each voxel that belongs to a nested entity, the uncertainty was set to 0 when the probability was higher than the corresponding high threshold. For each voxel that belongs to the background, the uncertainty was set to 0 when the maximum probability was lower than the low threshold. Such a method enabled the adjustment of the uncertainty of nested entities and the background independently.

\subsubsection{Method-5: Team Radiomics\_MIU \citep{RadiomicsMIU}}
The method used an ensemble of three different CNNs \citep{P-Net,U-Net,ResINC} for segmentation. Different models were trained for three different tumor entities (i.e., WT, TC, and ET segmentation). Three model ensembles were used, i.e., a total of nine models were trained for the task. Averaging various probabilities is one of the best and most effective ways to get a prediction of the ensemble model in classification. The uncertainty was estimated using the concept of entropy to represent voxel-wise variance and diversity information. The resulting uncertainty values were scaled to lie between 0 and 100.

\subsubsection{Method-6: Team Med\_vision \citep{MedVision}}
The method proposed self-ensemble-resUNet. The output softmax probabilities ($y_\text{pred}$) were inverted and normalized between 0-100 to obtain the uncertainty maps ($U_\textbf{pred}$): $U_\text{pred} = 100 \cdot (1 - y_\text{pred})$

\subsubsection{Method-7: Team Jaguars \citep{Jaguars}}
The method used an ensemble of a total of 7 U-Net type models. The output probabilities of each model were averaged for each label in each voxel to obtain a new probability for the ensemble. Since the model makes a binary classification of each voxel, the highest uncertainty corresponds with a probability of 0.5. Then the normalized entropy was used to get an uncertainty measure of the prediction for each voxel: $$ H = \sum_{c\in C} \frac{p_c  \cdot  \log (p_c)}{log(|C|)} \in [0,1], $$ where $p_c$ is the sigmoid output average probability of class c and C is the set of classes, (C = \{0,1\} in this case). These values were multiplied by 100 to normalize it between 0 and 100.

\subsubsection{Method-8: Team UmU \citep{UmU}}
The method proposes a Multi-Decoder Cascaded Network to predict the probability of the three tumor entities. An uncertainty score, $u^r_{i,j,k}$, at voxel $(i, j, k)$ was defined by:  $$u^r_{i,j,k} = \begin{cases} 200 \cdot (1-p^r_{i,j,k}), & \text{if} ~  p^r_{i,j,k} \ge 0.5 \\ 200  \cdot  p^r_{i,j,k}, & \text{if} ~ p^r_{i,j,k} < 0.5 \end{cases} $$ where $u^r_{i,j,k} \in [0, 100]^{|R|}$ and $p^r_{i,j,k} \in [0, 1]^{|R|}$ are the uncertainty score map and probability map, respectively. Here, $r\in R$, where $R$ is the set of tumor entities, i.e. WT, TC, and ET. 

\subsubsection{Method-9: Team LMB \citep{LMB}}
The method used a V-net \citep{VNet} architecture. A combination of test-time-dropout and test-time-augmentation was used for uncertainty estimation. In particular, the same input was passed through the network 20 times with random dropout and random data augmentation. The uncertainty map was estimated with the variance for each sub-region independently. Let $Y^i = {y_1^i, y_2^i, ..., y_B^i}$ be the vector that represents predicted labels for the $i^{th}$ voxel. The voxel-wise uncertainty map, for each tumor entity (WT,TC,ET), was obtained as the variance: $$ \text{var} = \frac{1}{B} \sum_{b=1}^B (y_b^i - y_{\text{mean}}^i)^2, $$ where $y_{\text{mean}}^i$ represents the mean prediction across $b$ samples.

\subsubsection{Method-10: Team Matukituki \citep{Matukituki}}
A multisequence 2D Dense-UNet segmentation model was trained. The final layer of this model is a four-channel soft-max layer representing the labels 'no tumor', 'edema', 'necrosis', and 'ET'. Uncertainty values were obtained from the final layer of the segmentation model for each label as follows: For WT, initial uncertainty values were obtained by adding the voxel-wise soft-max values of 'edema + necrosis + ET'. The initial uncertainty values for TC were the voxel-wise sum of 'necrosis + ET'. The initial uncertainty of the ET was the values of the voxel-wise soft-max channel representing ET. For all labels, the initial uncertainty values were clipped between 0 and 1. They were then modified according to the function: uncertainty = (1 – initial uncertainty) x 100. Finally, uncertainty values of 99 were changed to 100.

\subsubsection{Method-11: Team QTIM \citep{QTIM}}
The method used an ensemble of five networks to estimate voxel-wise segmentation uncertainty. Mirror axis-flipped inputs were passed through all models in the ensemble, resulting in 40 predictions per entity. These predictions were combined by directly averaging the model logits, denoted as $l_x$. A voxel with high predictive uncertainty will have $|l_x| \approx 0$, whereas a voxel with high predictive certainty will have $ |l_x| \gg 5$. To explicitly quantify uncertainty (U) in the range 0 (maximally certain) to 100 (maximally uncertain), the following formula is used: 
\begin{equation*}
    U_x = 
    \begin{cases} 
    200 \cdot \sigma(l_x) & \text{if } 0 \le \sigma(l_x) < 0.5 \\
    200  \cdot  (1 - \sigma(l_x)) & \text{otherwise} 
    \end{cases}
\end{equation*}
where the $\sigma$ function converts the ensembled logits to probabilities.

\subsubsection{Method-12: Team Nico@LRDE}
A cascade of two 3D U-Net type networks was employed for the task of brain tumor segmentation and its associated uncertainty estimation. The first network was trained for the brain tumor segmentation task. The second network was trained to predict where the segmentation network made wrong predictions. Here, the ground truth for training this network was generated as follows: the ground truth was considered ones (present) at voxels where the segmentation network was wrong, and it was considered as zeros (absent) at voxels where the segmentation network was correct. This way, the uncertainty networks learn to return zeros where the segmentation network is generally accurate and values next to one where the segmentation networks will have issues correctly predicting the segmentation ground truth. The output of the uncertainty estimation network (second network) was normalized between 0-100.

\section{Analysis}
This section presents the complete analyses and evaluation of teams that participated in the QU-BraTS 2020 challenge. Section \ref{RankScheme} provides the description of the evaluation and ranking strategy followed during the QU-BraTS 2020 challenge. Section \ref{sec:overallrank} provides the overall ranking results (accounting for all tumor entities) according to which the winning teams were announced at the challenge (Figure \ref{fig:QUBraTS_2020_ranking}). We also compare their ranking on the segmentation task in the same section. Then, Section \ref{sec:tsrank} provides the ranked order of the participating teams according to the individual tumor entities (Figure \ref{fig:QUBraTS_2020_ranking_WT}-\ref{fig:QUBraTS_2020_ranking_ET}), followed by our ablation study (in Section \ref{sec:ablation}) on the scores incorporated in the general score (Equation \ref{Equation-score}) (Figure \ref{fig:DICE_BraTS_2020_ranking}-\ref{fig:DICE_FTN_BraTS_2020_ranking}). Table \ref{tab:rankTable} encapsulates a summary of the ranked order of the participating teams for all this analysis. Finally, Section \ref{QualAnalysis} provides qualitative results highlighting the effect of uncertainty thresholding filtering for all participating teams. .

\begin{table}[t]
\caption{Summary of team ranking for different analyses performed in this paper. We use the ranking scheme described in Section:\ref{RankScheme} to rank different teams. ``QU-BraTS Ranking'' column depicts the actual team ranking for all participating teams in QU-BraTS 2020 challenge (Section \ref{sec:overallrank}). In the "Segmentation Ranking" column, we also report segmentation ranking for all teams that participated in the QU-BraTS challenge. The segmentation ranking is across 78 teams that participated in the segmentation task during BraTS 2020. In three columns under "Ranking based on Individual Tumor Entities" (Section \ref{sec:tsrank}), we provide a team ranking based only on one of the three tumor entities. Similarly, we also report the team ranking based on the ablation study of our developed score in the last three columns of "Ranking Based on Ablation Study" (Section \ref{sec:ablation}). For each type of ranking, the total number of provided ranks (given in the bracket) varies, as we provide the same rank for teams that do not have a significant statistical difference between their performance (Section  \ref{RankScheme}).}

\resizebox{\textwidth}{!}{%
\begin{tabular}{|l|cc|cccccc|}
\hline
\multirow{3}{*}{\textbf{Teams}} & \multicolumn{2}{c|}{\multirow{2}{*}{\textbf{\begin{tabular}[c]{@{}c@{}}Challenge Ranking\\ (Section 5.2.1)\end{tabular}}}}                                                     & \multicolumn{6}{c|}{\textbf{Variations}}                                                                                                                                                                                                                                                                                                                                                                                                                                                                            \\ \cline{4-9} 
                                & \multicolumn{2}{c|}{}                                                                                                                                                         & \multicolumn{3}{c|}{\textbf{\begin{tabular}[c]{@{}c@{}}Ranking Based on \\ Individual Tumor Entities\\ (Section 5.2.2)\end{tabular}}}                                                                                                                                                & \multicolumn{3}{c|}{\textbf{\begin{tabular}[c]{@{}c@{}}Ranking Based on \\ Ablation Study\\ (Section 5.2.3)\end{tabular}}}                                                                                                   \\ \cline{2-9} 
                                & \multicolumn{1}{c|}{\textbf{\begin{tabular}[c]{@{}c@{}}QU-BraTS \\ Ranking (9)\end{tabular}}} & \textbf{\begin{tabular}[c]{@{}c@{}}Segmentation \\ Ranking (18)\end{tabular}} & \multicolumn{1}{c|}{\textbf{\begin{tabular}[c]{@{}c@{}}Whole \\ Tumor (13)\end{tabular}}} & \multicolumn{1}{c|}{\textbf{\begin{tabular}[c]{@{}c@{}}Tumor \\ Core (11)\end{tabular}}} & \multicolumn{1}{c|}{\textbf{\begin{tabular}[c]{@{}c@{}}Enhancing \\ Tumor (11)\end{tabular}}} & \multicolumn{1}{c|}{\textbf{DSC AUC (10)}} & \multicolumn{1}{c|}{\textbf{\begin{tabular}[c]{@{}c@{}}DSC AUC and \\ FTP AUC (9)\end{tabular}}} & \textbf{\begin{tabular}[c]{@{}c@{}}DSC AUC and \\ FTN AUC (12)\end{tabular}} \\ \hline
\textbf{SCAN}                   & \multicolumn{1}{c|}{1}                                                                        & 4                                                                             & \multicolumn{1}{c|}{1}                                                                    & \multicolumn{1}{c|}{1}                                                                   & \multicolumn{1}{c|}{1}                                                                        & \multicolumn{1}{c|}{6}                     & \multicolumn{1}{c|}{2}                                                                           & 4                                                                            \\ \hline
\textbf{UmU}                    & \multicolumn{1}{c|}{2}                                                                        & 7                                                                             & \multicolumn{1}{c|}{3}                                                                    & \multicolumn{1}{c|}{2}                                                                   & \multicolumn{1}{c|}{2}                                                                        & \multicolumn{1}{c|}{4}                     & \multicolumn{1}{c|}{3}                                                                           & 3                                                                            \\ \hline
\textbf{DSI\_Med}               & \multicolumn{1}{c|}{2}                                                                        & 13                                                                            & \multicolumn{1}{c|}{2}                                                                    & \multicolumn{1}{c|}{2}                                                                   & \multicolumn{1}{c|}{3}                                                                        & \multicolumn{1}{c|}{9}                     & \multicolumn{1}{c|}{3}                                                                           & 7                                                                            \\ \hline
\textbf{QTIM}                   & \multicolumn{1}{c|}{3}                                                                        & 7                                                                             & \multicolumn{1}{c|}{4}                                                                    & \multicolumn{1}{c|}{2}                                                                   & \multicolumn{1}{c|}{3}                                                                        & \multicolumn{1}{c|}{3}                     & \multicolumn{1}{c|}{4}                                                                           & 2                                                                            \\ \hline
\textbf{Uniandes}               & \multicolumn{1}{c|}{4}                                                                        & 15                                                                            & \multicolumn{1}{c|}{5}                                                                    & \multicolumn{1}{c|}{3}                                                                   & \multicolumn{1}{c|}{4}                                                                        & \multicolumn{1}{c|}{8}                     & \multicolumn{1}{c|}{5}                                                                           & 6                                                                            \\ \hline
\textbf{nsu\_btr}               & \multicolumn{1}{c|}{5}                                                                        & 13                                                                            & \multicolumn{1}{c|}{10}                                                                   & \multicolumn{1}{c|}{8}                                                                   & \multicolumn{1}{c|}{10}                                                                       & \multicolumn{1}{c|}{1}                     & \multicolumn{1}{c|}{4}                                                                           & 9                                                                            \\ \hline
\textbf{LMB}                    & \multicolumn{1}{c|}{5}                                                                        & 20                                                                            & \multicolumn{1}{c|}{8}                                                                    & \multicolumn{1}{c|}{4}                                                                   & \multicolumn{1}{c|}{3}                                                                        & \multicolumn{1}{c|}{10}                    & \multicolumn{1}{c|}{7}                                                                           & 8                                                                            \\ \hline
\textbf{radiomics\_miu}         & \multicolumn{1}{c|}{6}                                                                        & 13                                                                            & \multicolumn{1}{c|}{7}                                                                    & \multicolumn{1}{c|}{5}                                                                   & \multicolumn{1}{c|}{5}                                                                        & \multicolumn{1}{c|}{2}                     & \multicolumn{1}{c|}{8}                                                                           & 3                                                                            \\ \hline
\textbf{Nico@LRDE}              & \multicolumn{1}{c|}{6}                                                                        & 18                                                                            & \multicolumn{1}{c|}{6}                                                                    & \multicolumn{1}{c|}{6}                                                                   & \multicolumn{1}{c|}{6}                                                                        & \multicolumn{1}{c|}{7}                     & \multicolumn{1}{c|}{9}                                                                           & 5                                                                            \\ \hline
\textbf{Jaguars}                & \multicolumn{1}{c|}{6}                                                                        & 13                                                                            & \multicolumn{1}{c|}{5}                                                                    & \multicolumn{1}{c|}{6}                                                                   & \multicolumn{1}{c|}{6}                                                                        & \multicolumn{1}{c|}{2}                     & \multicolumn{1}{c|}{8}                                                                           & 3                                                                            \\ \hline
\textbf{Team\_Alpaca}           & \multicolumn{1}{c|}{7}                                                                        & 10                                                                            & \multicolumn{1}{c|}{9}                                                                    & \multicolumn{1}{c|}{7}                                                                   & \multicolumn{1}{c|}{7}                                                                        & \multicolumn{1}{c|}{2}                     & \multicolumn{1}{c|}{1}                                                                           & 1                                                                            \\ \hline
\textbf{Matukituki}             & \multicolumn{1}{c|}{8}                                                                        & 19                                                                            & \multicolumn{1}{c|}{11}                                                                   & \multicolumn{1}{c|}{9}                                                                   & \multicolumn{1}{c|}{9}                                                                        & \multicolumn{1}{c|}{7}                     & \multicolumn{1}{c|}{4}                                                                           & 12                                                                           \\ \hline
\textbf{Frankenstein}           & \multicolumn{1}{c|}{9}                                                                        & 18                                                                            & \multicolumn{1}{c|}{13}                                                                   & \multicolumn{1}{c|}{11}                                                                  & \multicolumn{1}{c|}{8}                                                                        & \multicolumn{1}{c|}{6}                     & \multicolumn{1}{c|}{6}                                                                           & 11                                                                           \\ \hline
\textbf{med\_vision}            & \multicolumn{1}{c|}{9}                                                                        & 14                                                                            & \multicolumn{1}{c|}{12}                                                                   & \multicolumn{1}{c|}{10}                                                                  & \multicolumn{1}{c|}{11}                                                                       & \multicolumn{1}{c|}{5}                     & \multicolumn{1}{c|}{7}                                                                           & 10                                                                           \\ \hline
\end{tabular}%
}
\label{tab:rankTable}
\end{table}

\subsection{Ranking Scheme: BraTS 2020 challenge on uncertainty quantification (QU-BraTS)}
\label{RankScheme}
  
The ranking scheme used during the challenge comprised the ranking of each team relative to its competitors for each testing subject, for each evaluated tumor entity (i.e., ET, TC, WT) using the overall score (Equation~\ref{Equation-score}). This ranking scheme led to each team being ranked for 166 subjects for three regions, resulting in 498 individual rankings. For each team, first, the individual ranking for each patient was calculated by adding ranks across each region. This ranking is referred to as the Cumulative Ranking Score (CRS). For each team, the Normalized Ranking Score (NRS) was also calculated for each patient by dividing their CRS by the total number of participating teams and the total number of regions. The NRS is in the range of 0-1 for each patient. The final ranking score (FRS) was calculated by averaging the cumulative rank across all patients for each participating team. Other challenges, such as the Ischemic Stroke Lesion Segmentation Challenge (ISLES - \url{http://www.isles-challenge.org/}) \citep{ISLES}, use a similar ranking scheme. \\

Following the BraTS challenge, further permutation testing was done to determine the statistical significance of the relative rankings between each pair of teams. This permutation testing would reflect differences in performance that exceeded those that might be expected by chance. Specifically, for each team, given a list of observed patient-level Cumulative Ranks, i.e., the actual ranking described above, for each pair of teams, repeated random permutations (i.e., 100,000 times) of the Cumulative Ranks for each subject were performed. The difference in the FRS between this pair of teams was calculated for each permutation. The proportion of times the difference in FRS calculated using randomly permuted data exceeded the observed difference in FRS (i.e., using the actual data) indicated the statistical significance of their relative rankings as a p-value. Teams that do not have a statistically significant difference in their FRS have similar respective ranks (group) on the leaderboard\footnote{Throughout the paper, we report any p-value less than 0.05 as the threshold for statistically significant differences.}.

\subsection{Team Ranking} \label{sec:teamrank}
This section reports the final rankings of all participating teams on BraTS 2020 test dataset. \\

\subsubsection{Overall Ranking Results} \label{sec:overallrank}

\begin{figure*}[t]
\centering
\includegraphics[width=1.00\textwidth]{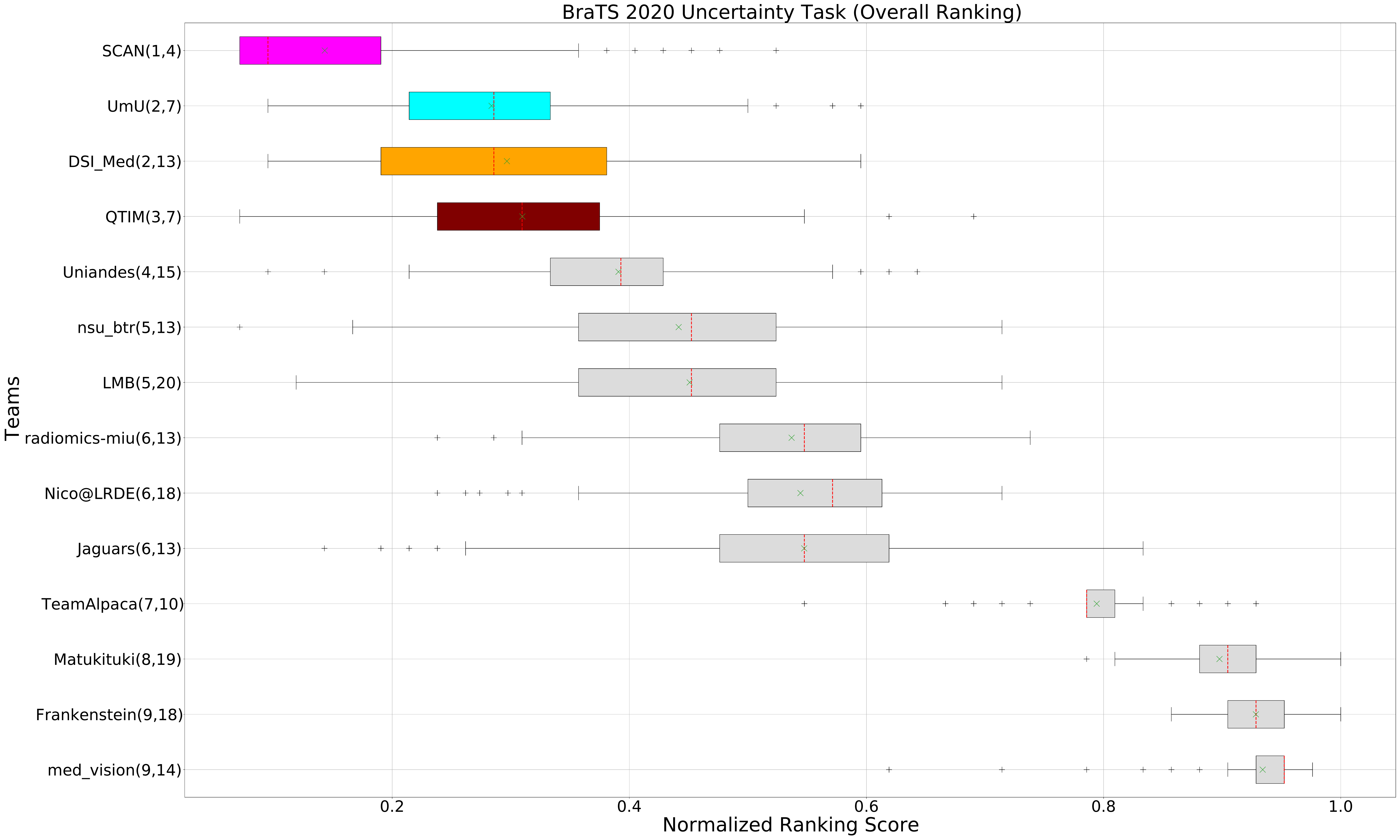}
\center \caption{QU-BraTS 2020 boxplots of Normalized Ranking Score (NRS) across patients for all participants on the BraTS 2020 test set (lower is better). Boxplots for the top four performing teams are visualized using Pink (\textit{Team SCAN}), orange (\textit{Team DSI\_Med}), Cyan (\textit{Team UmU}), and Maroon (\textit{Team QTIM}) colour. Box plots for the remaining teams use gray colour. Y-axis shows the name of each team and their respective uncertainty task ranking, followed by their segmentation task ranking. There was no statistically significant difference between per patient ranking of teams ranked at the same position. Teams which has different ranks had statistically significant differences in their per-patient ranking.}
\label{fig:QUBraTS_2020_ranking}
\end{figure*}

Figure~\ref{fig:QUBraTS_2020_ranking} (and QU-BraTS ranking column in Table~\ref{tab:rankTable}) provides a relative ranking for each team\footnote{Box plot depicting performance of each team for four different scores - DICE\_AUC, FTP\_RATIO\_AUC, FTN\_RATIO\_AUC, SCORE, for three different tumor entities - WT, TC, ET, is given in Appendix A.}. We can see that \textit{Team SCAN} comfortably outperforms all other methods and achieves the first rank in the challenge. Their Normalized Ranking Score (NRS) across all patients was $\sim 0.14$, while the NRS (across all patients) for the teams which achieved rank 2 (\textit{Team UmU} and \textit{Team DSI\_Med}) was $\sim 0.28$. There was no statistically significant difference between \textit{Team UmU} and \textit{Team DSI\_Med}. Thus both teams were ranked at position 2 on the challenge leaderboard. \textit{Team QTIM} ranked 3rd in the challenge leaderboard and achieved marginally (though statistically significant) lower performance compared to Rank-2 teams (average NRS of $\sim 0.31$ compared to average NRS of $\sim 0.28$). \\

We also report the relative segmentation ranking of each team participating in the uncertainty challenge. The reported segmentation task ranking is across 78 teams that participated in the segmentation task. From Figure Figure~\ref{fig:QUBraTS_2020_ranking} (and Segmentation Ranking column in Table~\ref{tab:rankTable}), we can observe that while the \textit{Team SCAN} (pink colour) achieves a higher ranking (Rank-4) than other teams in the segmentation task, the segmentation task ranking and the uncertainty task (QU-BraTS challenge) ranking are not the same. This is visible for \textit{Team UmU} and \textit{Team QTIM}, as both achieved a similar ranking (rank-7) in the segmentation task of BraTS 2020; while \textit{Team UmU} was ranked second in the uncertainty task, \textit{Team QTIM} was ranked third. Similarly, we can observe that three teams that achieved Rank-13 in the segmentation task (\textit{Team DSI\_Med}, \textit{Team nsu\_btr}, and \textit{radiomics-miu}) were ranked differently in the uncertainty evaluation task (Rank-2, Rank-5, and Rank-6, respectively). The difference in ranking across both tasks shows that performing well on the segmentation task does not guarantee good performance on the uncertainty evaluation task, and both tasks are complementary. \\ 

\begin{figure*}[t]
\centering
\includegraphics[width=1.00\textwidth]{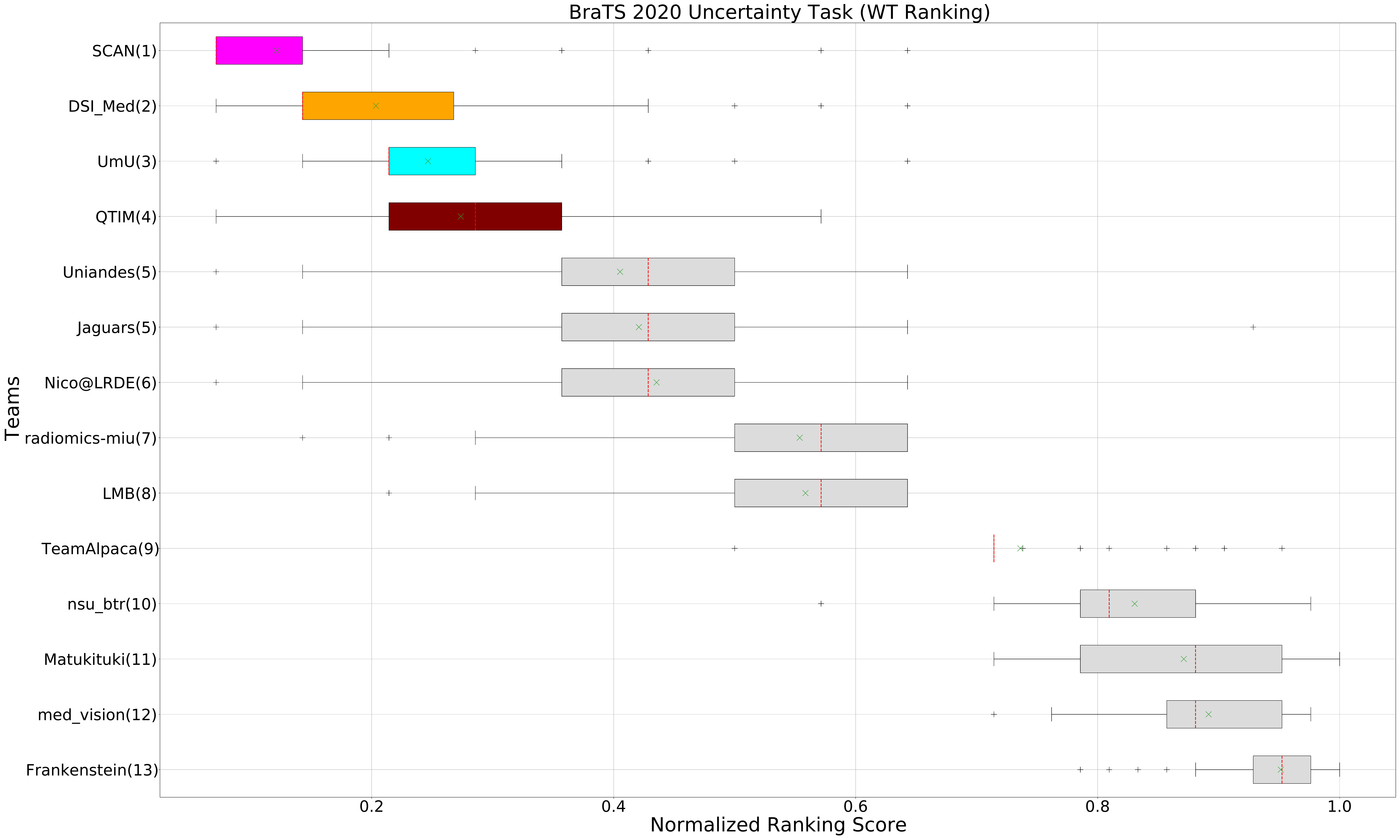} 
\center \caption{QU-BraTS 2020 boxplots of Normalized Ranking Score (NRS) across patients for all participants on the BraTS 2020 test set test set only for Whole Tumor (lower is better). Boxplots for the top four performing teams (in the final ranking - Figure~\ref{fig:QUBraTS_2020_ranking}) are visualized using Pink (\textit{Team SCAN}), orange (\textit{Team DSI\_Med}), Cyan (\textit{Team UmU}), and Maroon (\textit{Team QTIM}) colour. Box plots for the remaining teams use gray colour. Y-axis shows the name of each team and their respective uncertainty task ranking, followed by their segmentation task ranking. There was no statistically significant difference between per patient ranking of teams ranked at the same position. Teams which has different ranks had statistically significant differences in their per-patient ranking.}
\label{fig:QUBraTS_2020_ranking_WT}
\end{figure*}

\begin{figure*}[t]
\centering
\includegraphics[width=1.00\textwidth]{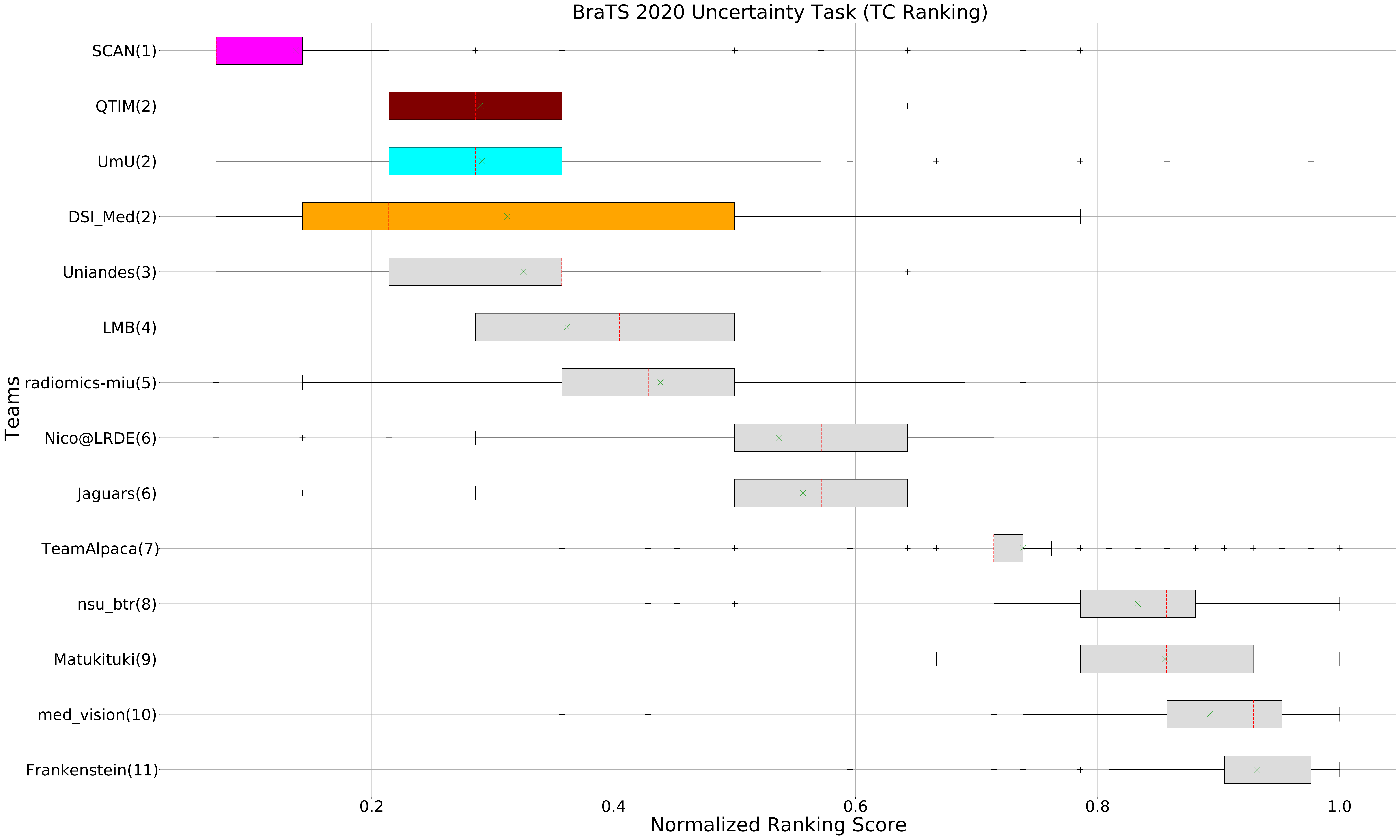} 
\center \caption{QU-BraTS 2020 boxplots of Normalized Ranking Score (NRS) across patients for all participants on the BraTS 2020 test set test set only for Tumor Core (lower is better). Boxplots for the top four performing teams (in the final ranking - Figure~\ref{fig:QUBraTS_2020_ranking}) are visualized using Pink (\textit{Team SCAN}), orange (\textit{Team DSI\_Med}), Cyan (\textit{Team UmU}), and Maroon (\textit{Team QTIM}) colour. Box plots for the remaining teams use gray colour. Y-axis shows the name of each team and their respective uncertainty task ranking, followed by their segmentation task ranking. There was no statistically significant difference between per patient ranking of teams ranked at the same position. Teams which has different ranks had statistically significant differences in their per-patient ranking.}
\label{fig:QUBraTS_2020_ranking_TC}
\end{figure*}

\begin{figure*}[t]
\centering
\includegraphics[width=1.00\textwidth]{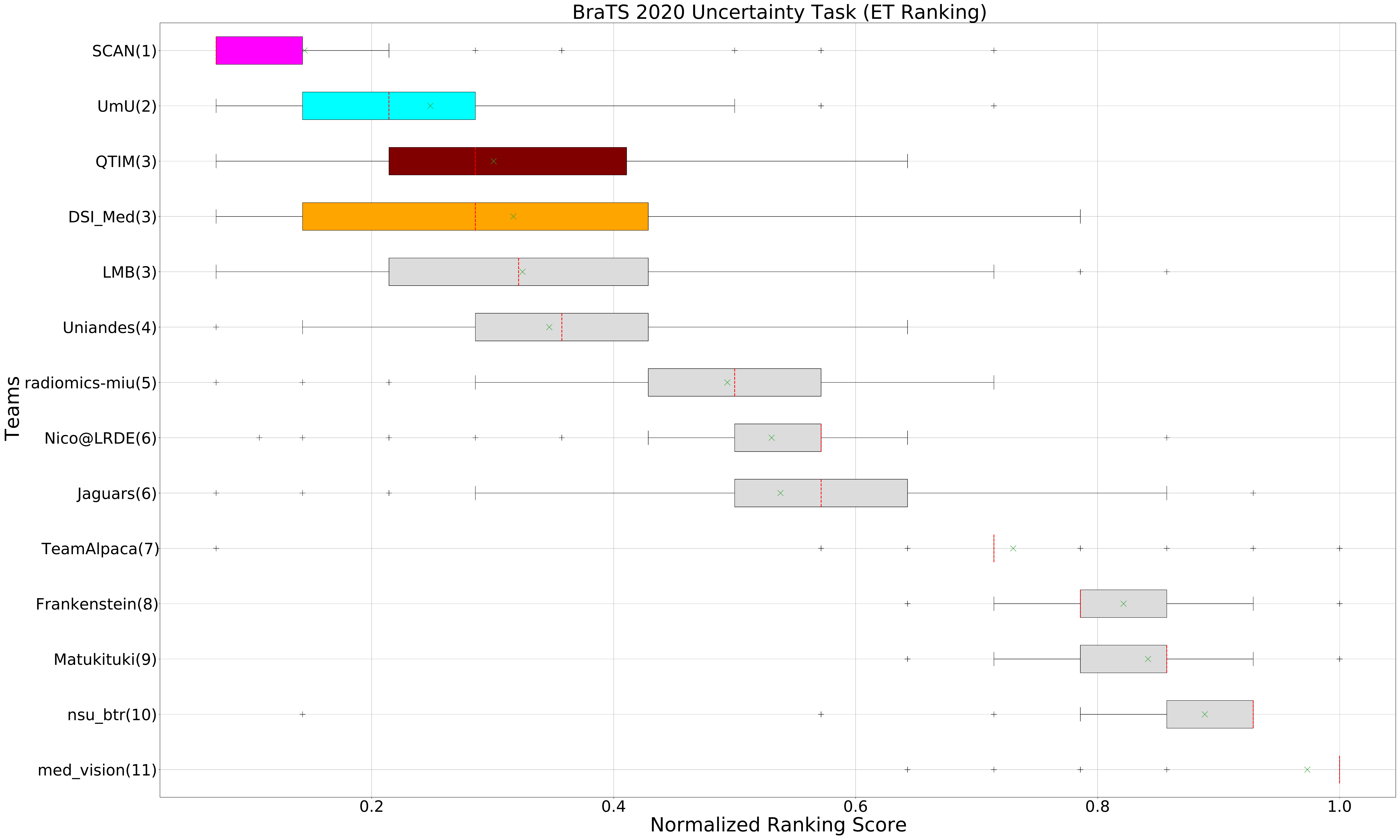} 
\center \caption{QU-BraTS 2020 boxplots of Normalized Ranking Score (NRS) across patients for all participants on the BraTS 2020 test set test set only for Enhancing Tumor (lower is better). Boxplots for the top four performing teams (in the final ranking - Figure~\ref{fig:QUBraTS_2020_ranking}) are visualized using Pink (\textit{Team SCAN}), orange (\textit{Team DSI\_Med}), Cyan (\textit{Team UmU}), and Maroon (\textit{Team QTIM}) colour. Box plots for the remaining teams use gray colour. Y-axis shows the name of each team and their respective uncertainty task ranking, followed by their segmentation task ranking. There was no statistically significant difference between per patient ranking of teams ranked at the same position. Teams which has different ranks had statistically significant differences in their per-patient ranking.}
\label{fig:QUBraTS_2020_ranking_ET}
\end{figure*}

\subsubsection{Team Ranking for individual tumor entities} \label{sec:tsrank}
The BraTS challenge involves three separate tumor entities (WT, TC, and ET). The segmentation performance across these entities varies, as reported in the previous BraTS challenge reports \citep{BraTS2015, BraTS2018}. Specifically, the BraTS challenge reports good $DSC$ across different teams for the WT segmentation task, while the performance for the ET segmentation task is relatively lower. The performance gap between different tumor entities can hinder the clinical adaptation of the segmentation algorithms. The main goal for developing the uncertainty evaluation scores is to make algorithms more useful for clinical adaptation. Keeping this in mind, we further report the raking of each participating team according to the score (Equation~\ref{Equation-score}) calculated for each tumor entity in Figure~\ref{fig:QUBraTS_2020_ranking_WT}, Figure~\ref{fig:QUBraTS_2020_ranking_TC}, and Figure~\ref{fig:QUBraTS_2020_ranking_ET}. \\

When teams are ranked only based on their WT scores (Figure~\ref{fig:QUBraTS_2020_ranking_WT} and Whole Tumor column in Table~\ref{tab:rankTable}), \textit{Team SCAN} still comfortably outperforms other teams similar to the original ranking (Figure~\ref{fig:QUBraTS_2020_ranking}). Unlike the original ranking scheme, \textit{Team DSI\_Med} ranks statistically significantly higher compared to \textit{Team UmU}. Similarly, from Figure~\ref{fig:QUBraTS_2020_ranking_TC} (and the Tumor Core column in Table~\ref{tab:rankTable}), we can observe that \textit{Team QTIM}, \textit{Team UmU}, and \textit{Team DSI\_Med} perform similarly without any statistically significant difference when ranked only based on their TC score as all teams are ranked at the same position. In Figure~\ref{fig:QUBraTS_2020_ranking_ET} (and the Enhancing Tumor column in Table~\ref{tab:rankTable}), \textit{Team UmU} achieves rank-2 with statistical significance compared to \textit{Team QTIM} and \textit{Team DSI\_Med}. We also observe no statistically significant difference between \textit{Team QTIM}, \textit{Team DSI\_Med}, and \textit{Team LMB}. \\

Overall, \textit{Team SCAN} comfortably ranks first for all tumor entities. \textit{Team UmU} ranks 3-2-2 for WT-TC-ET, while \textit{Team DSI\_Med} ranks 2-2-3 for WT-TC-ET. Both teams are ranked at position 2 when considering all tumor entities. The analysis shows that different teams achieve different ranks depending on the tumor entities, which shows that their performance differs across different tumor entities.  

\subsubsection{Ablation study on our score} \label{sec:ablation}
The overall score for uncertainty evaluation is calculated as a combination of three different AUCs as described in  Equation~\ref{Equation-score}. Section~\ref{UncMetric} described the  rationale behind the development of this score. As discussed in Section~\ref{UncMetric}, evaluating the task-dependent metric (in our case, $DSC$) as a function of filtered samples is critical, especially in the case of pathology segmentation, where there is a high class imbalance. We expect that, by filtering voxels with decrease in the uncertainty threshold, the performance on the remaining voxels measured using the task-dependent metric ($DSC$) should increase but not at the expense of filtering true positive or true negative voxels. The final score consists of the task-dependent metric and filtered true positive/negatives as a function of uncertainty thresholds. In this section, we perform an ablation study of different components of the final score ($DSC$, FTP, FTN). Our analysis reaffirms that only considering one or two components of the final score leads to a different ranking among participating teams. \\

\begin{figure*}[t]
\centering
\includegraphics[width=1.00\textwidth]{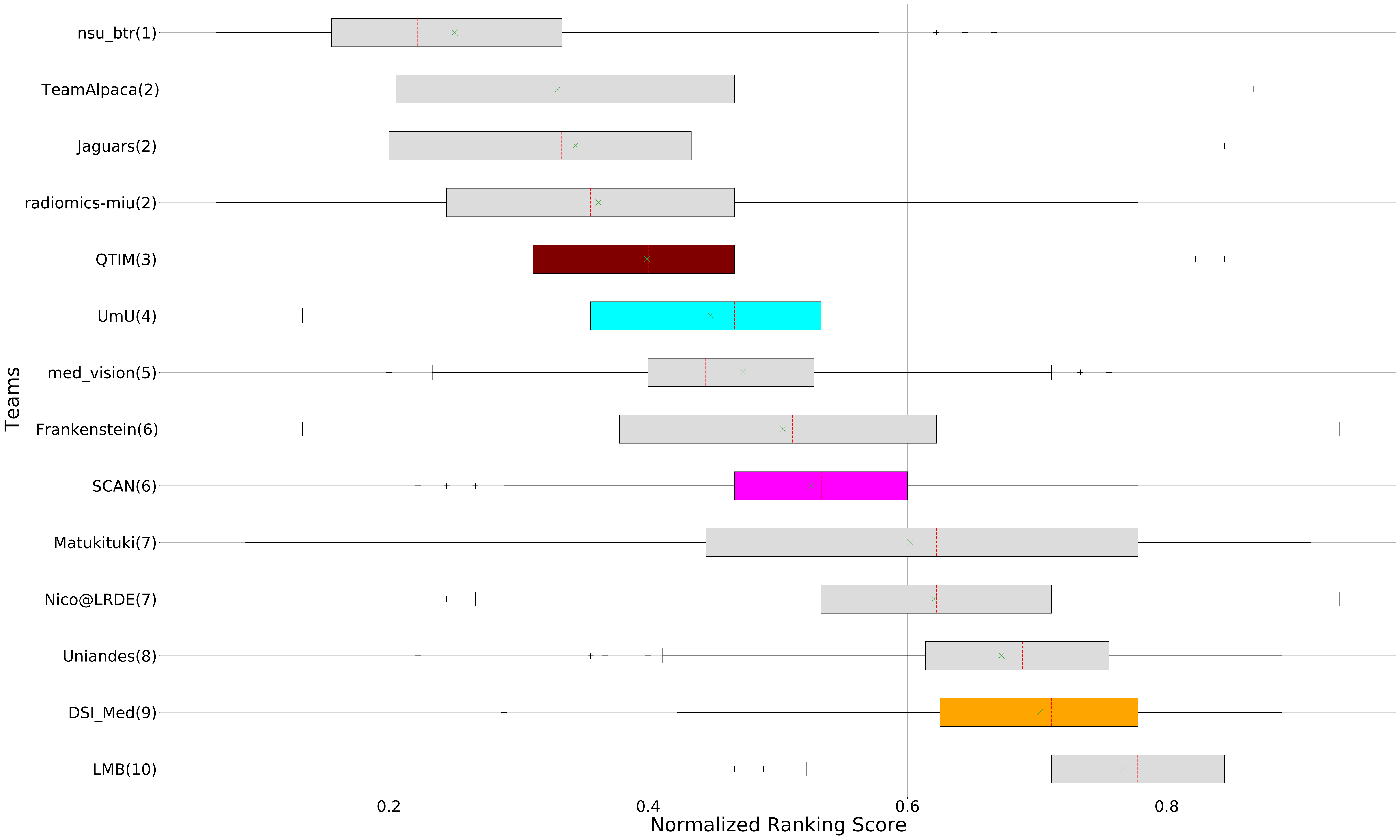} 
\center \caption{QU-BraTS 2020 boxplots of Normalized Ranking Score (NRS) across patients for all participants on the BraTS 2020 test set test set based only on DICE\_AUC score (lower is better). Boxplots for the top four performing teams (in the final ranking - Figure~\ref{fig:QUBraTS_2020_ranking}) are visualized using Pink (\textit{Team SCAN}), orange (\textit{Team DSI\_Med}), Cyan (\textit{Team UmU}), and Maroon (\textit{Team QTIM}) colour. Box plots for the remaining teams use gray colour. Y-axis shows the name of each team and their respective uncertainty task ranking, followed by their segmentation task ranking. There was no statistically significant difference between per patient ranking of teams ranked at the same position. Teams which has different ranks had statistically significant differences in their per-patient ranking. }
\label{fig:DICE_BraTS_2020_ranking}
\end{figure*}

\begin{figure*}[t]
\centering
\includegraphics[width=1.00\textwidth]{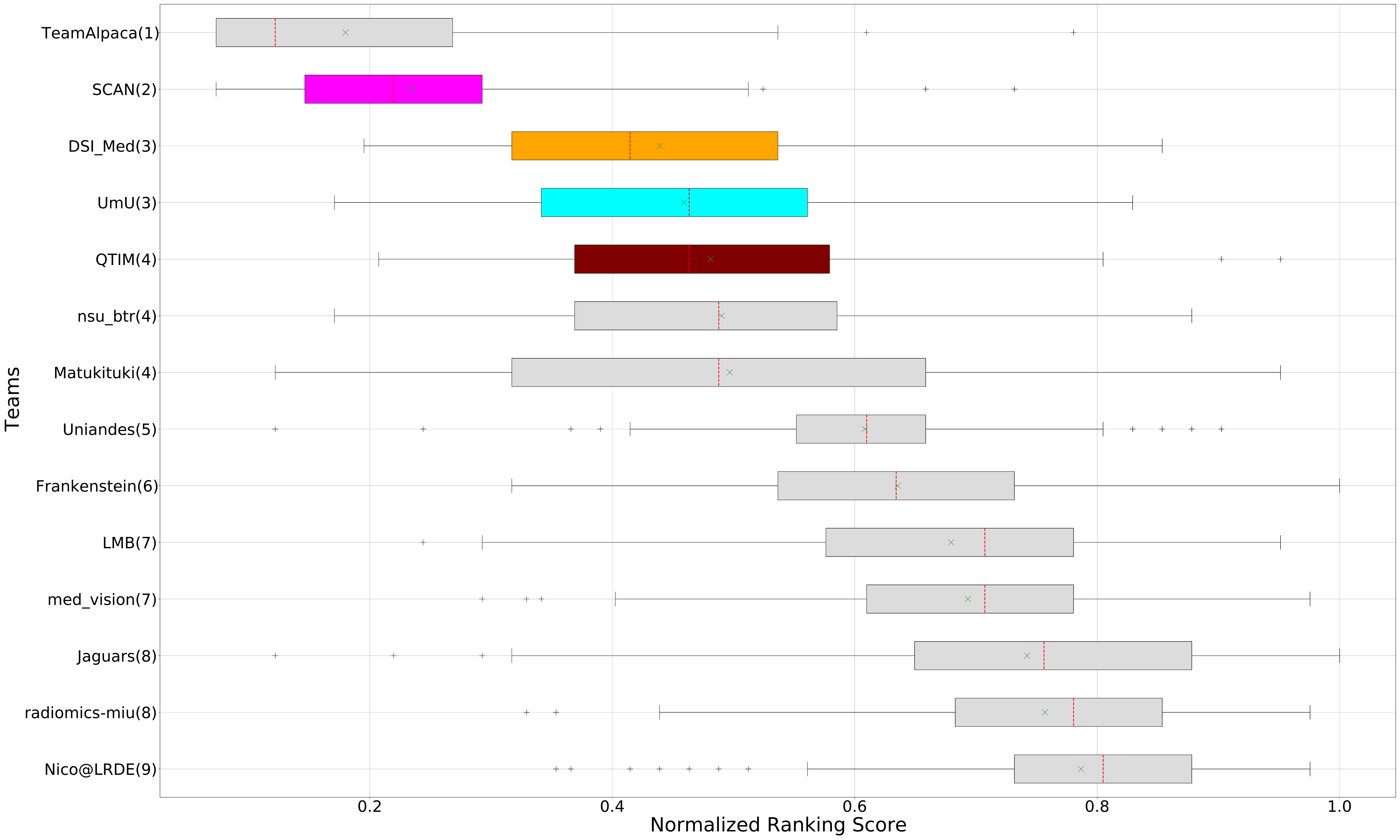}
\center \caption{QU-BraTS 2020 boxplots of Normalized Ranking Score (NRS) across patients for all participants on the BraTS 2020 test set test set based on a combination of DICE\_AUC score and FTP\_AUC score (lower is better). Boxplots for the top four performing teams (in the final ranking - Figure~\ref{fig:QUBraTS_2020_ranking}) are visualized using Pink (\textit{Team SCAN}), orange (\textit{Team DSI\_Med}), Cyan (\textit{Team UmU}), and Maroon (\textit{Team QTIM}) colour. Box plots for the remaining teams use gray colour. Y-axis shows the name of each team and their respective uncertainty task ranking, followed by their segmentation task ranking. There was no statistically significant difference between per patient ranking of teams ranked at the same position. Teams which has different ranks had statistically significant differences in their per-patient ranking.}
\label{fig:DICE_FTP_BraTS_2020_ranking}
\end{figure*}

\begin{figure*}[t]
\centering
\includegraphics[width=1.00\textwidth]{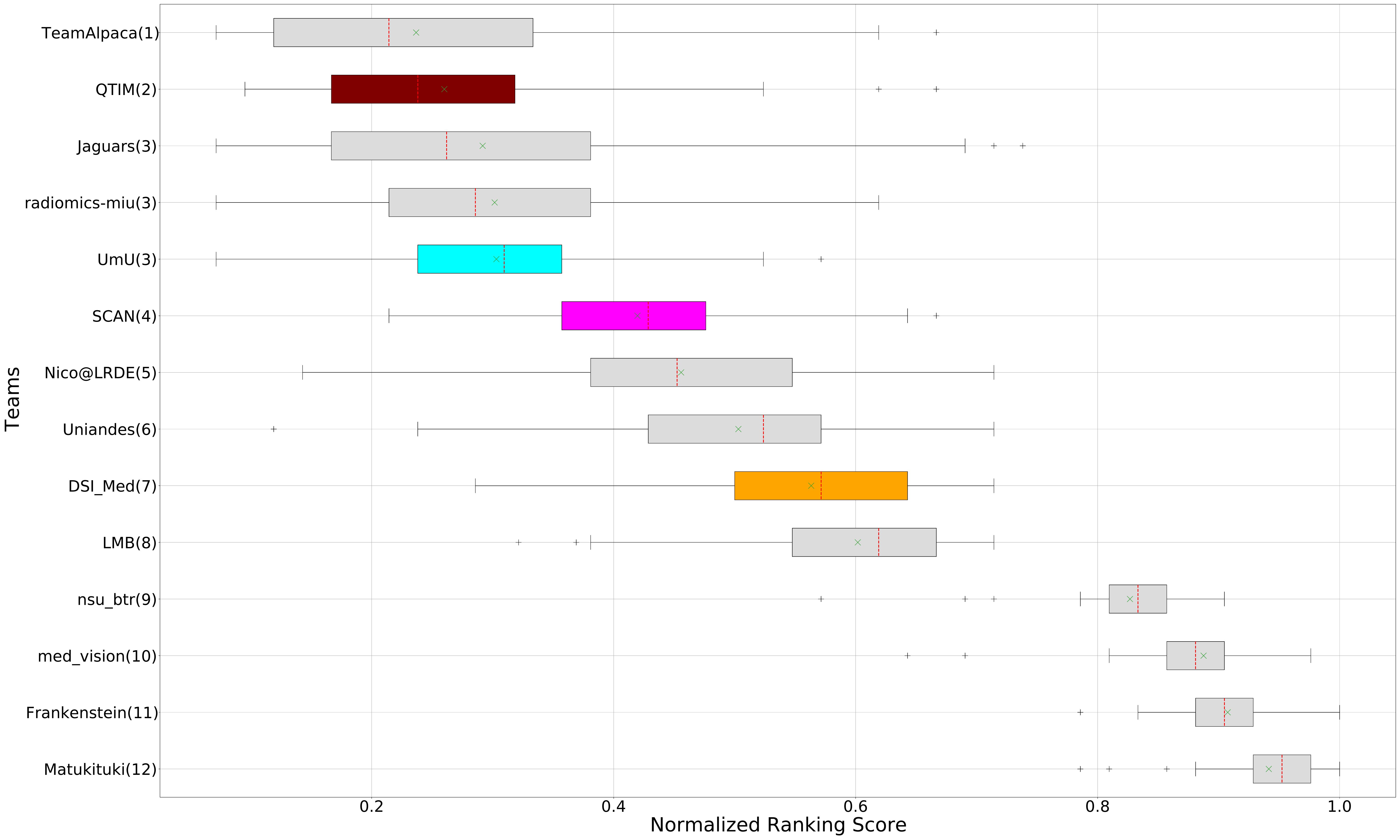}
\center \caption{QU-BraTS 2020 boxplots of Normalized Ranking Score (NRS) across patients for all participants on the BraTS 2020 test set test set based on a combination of DICE\_AUC score and FTN\_AUC score (lower is better). Boxplots for the top four performing teams (in the final ranking - Figure~\ref{fig:QUBraTS_2020_ranking}) are visualized using Pink (\textit{Team SCAN}), orange (\textit{Team DSI\_Med}), Cyan (\textit{Team UmU}), and Maroon (\textit{Team QTIM}) colour. Box plots for the remaining teams use gray colour. Y-axis shows the name of each team and their respective uncertainty task ranking, followed by their segmentation task ranking. There was no statistically significant difference between per patient ranking of teams ranked at the same position. Teams which has different ranks had statistically significant differences in their per-patient ranking.}
\label{fig:DICE_FTN_BraTS_2020_ranking}
\end{figure*}

\begin{figure*}[t]
\centering
\includegraphics[width=0.6\textwidth]{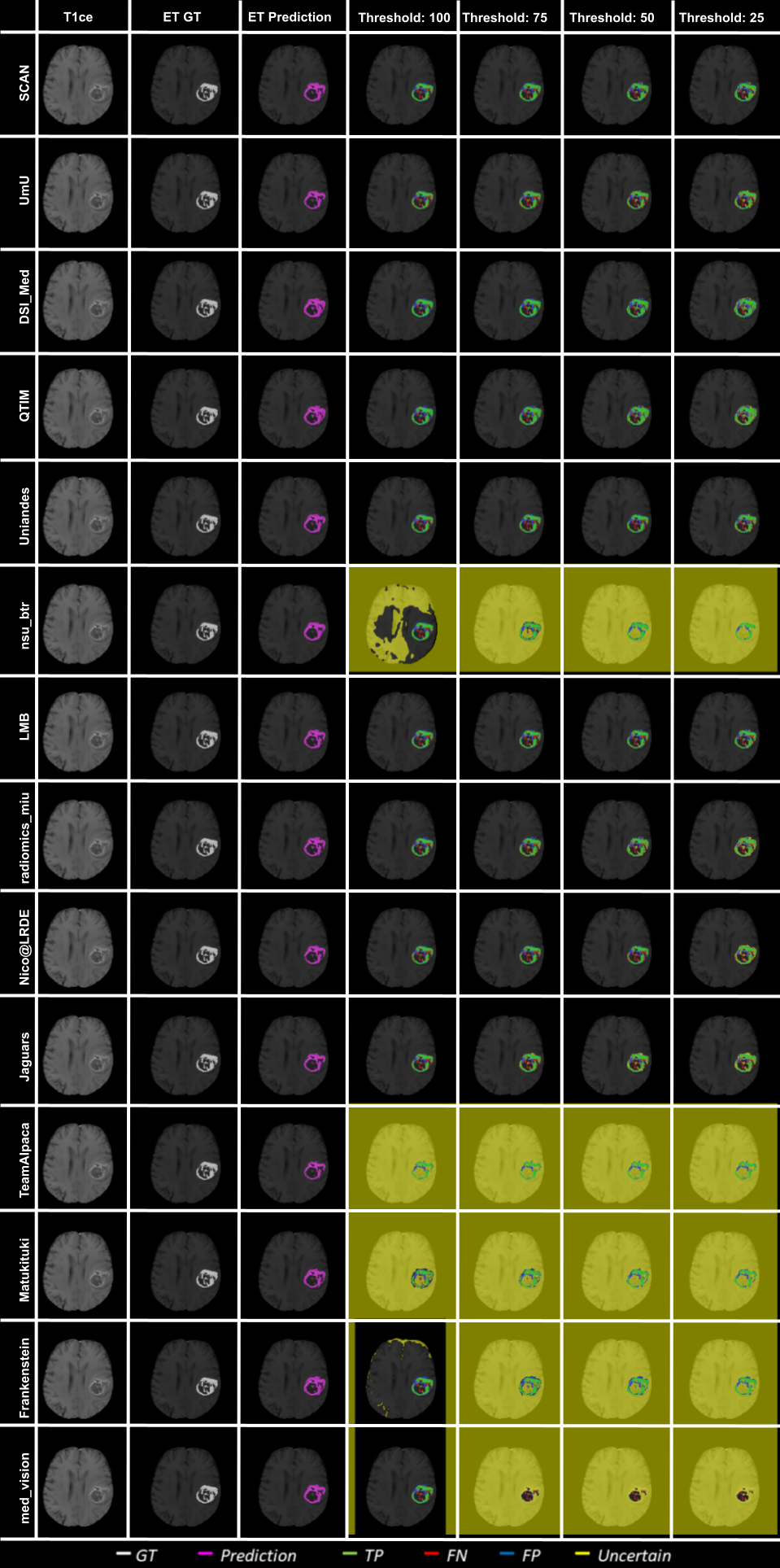} 
\center \caption{Effect of uncertainty thresholding on a BraTS 2020 test case for enhance tumor segmentation across different participating teams. (a) T1ce MRI (b) Ground Truth (c) Prediction (d) No filtering. Uncertainty Threshold = 100 (e) Uncertainty Threshold = 75 (f) Uncertainty Threshold = 50 (g) Uncertainty Threshold = 25.}
\label{fig:Qual_enhance_135_100}
\end{figure*}

\noindent \textbf{Ranking According to $DSC$ AUC:} The main component of any uncertainty evaluation score is the task dependent metric, in our case, $DSC$. Many previously proposed methods for various tasks only report the value of task dependent metrics at various uncertainty filtering thresholds -- For example, the AUC score for multiple sclerosis \cite{BUNet}. In Figure~\ref{fig:DICE_BraTS_2020_ranking} (and the $DSC$ AUC column in Table~\ref{tab:rankTable}), we rank participating teams according to their performance based on the AUC of $DSC$ vs. Uncertainty threshold. The figure shows that higher ranking teams in this ranking scheme (\textit{Team nsu\_btr}, \textit{Team Alpaca}, and \textit{Team Jaguars}) are different from those (\textit{Team SCAN}, \textit{Team UmU}, and \textit{Team DSI\_Med}) in the original ranking scheme (Figure~\ref{fig:QUBraTS_2020_ranking}).
A closer look at the higher ranking teams according to AUC of $DSC$ (Figure~\ref{fig:DICE_BraTS_2020_ranking}) reveals that teams like \textit{Team Alpaca} (Section~\ref{TeamAlpaca}) achieve a good score by using $100 - (100 \cdot \text{softmax}\_\text{confidence})$ as a proxy for uncertainty. 
Using softmax confidence in the foreground class (e.g. tumour subclass) as a direct proxy to uncertainty leads to all voxels belonging to the background class (i.e. healthy tissues) being marked as uncertain at a low uncertainty threshold. This would increase the burden in a system where we are asking clinicians to review all uncertain voxels (Figure~\ref{fig:Qual_enhance_135_100}). We observed that Team Alpaca used softmax confidence in the foreground class as a direct proxy to uncertainty.\\

\noindent \textbf{Ranking according to a combination of $DSC$ AUC and FTP or FTN AUC:} In the last section, we ranked teams according to their performance on the task-dependent evaluation metrics ($DSC$) at different uncertainty thresholds. As mentioned in Section~\ref{UncMetric}, ranking teams only based on their task-dependent evaluation metric rewards methods which filter out many positive predictions at low uncertainty thresholds to attain higher performance on the metric of interest. This would increase the burden in scenarios where clinical review is needed for all uncertain predictions. To alleviate the issue, teams are ranked according to a combination of (i) AUC score for $DSC$ and (ii) AUC for FTP or AUC for FTN. From Figure~\ref{fig:DICE_FTP_BraTS_2020_ranking} (and $DSC$ AUC and FTP AUC column in Table~\ref{tab:rankTable}), we can conclude that a combination of both DICE\_AUC and FTP\_AUC alone is insufficient. It still leads to \textit{Team Alpaca} ranked higher. As shown in Figure~\ref{fig:Qual_enhance_135_100}, \textit{Team Alpaca} marks all healthy-tissues (True Negative) voxels as uncertain, which reflects that the segmentation method is not confident in its prediction of healthy tissue. This is problematic as it would increase the burden in scenarios where we expect clinicians to review all uncertain predictions. We see a similar problem when teams are ranked only using a combination of DICE\_AUC and FTN\_AUC (Figure~\ref{fig:DICE_FTN_BraTS_2020_ranking} and $DSC$ AUC and FTN AUC column in Table~\ref{tab:rankTable}). \\

Analysis in the previous two sections highlights the necessity of combining all three AUCs to calculate the final score for ranking teams in the context of uncertainty quantification of the brain tumor segmentation task.

\begin{figure*}[t]
\centering
\includegraphics[width=0.6\textwidth]{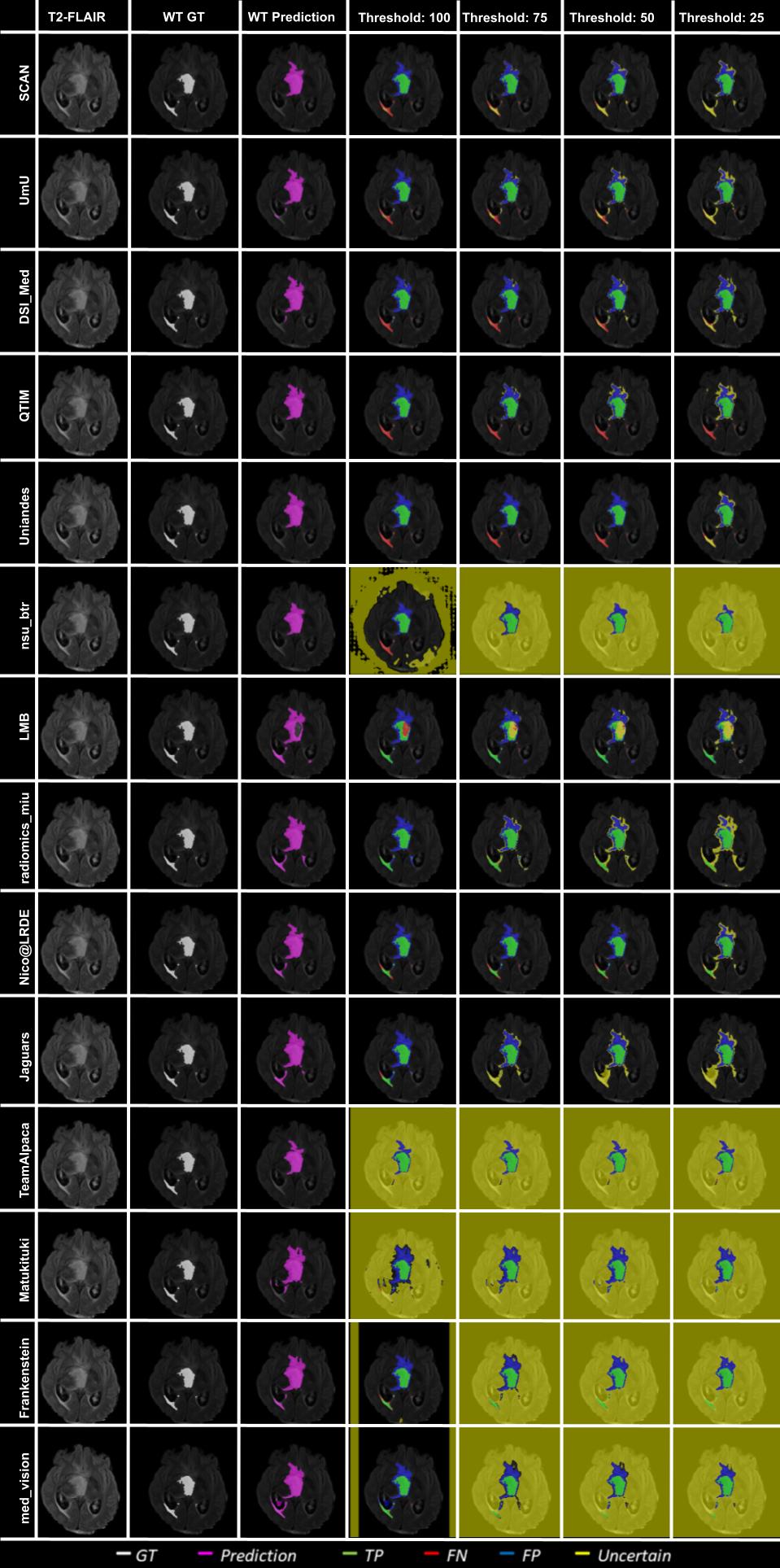}
\center \caption{Effect of uncertainty thresholding on a BraTS 2020 test case for whole tumor segmentation across different participating teams. (a) T2-FLAIR MRI (b) Ground Truth (c) Prediction (d) No filtering. Uncertainty Threshold = 100 (e) Uncertainty Threshold = 75 (f) Uncertainty Threshold = 50 (g) Uncertainty Threshold = 25.}
\label{fig:Qual_whole_017_066}
\end{figure*}

\begin{figure*}[t]
\centering
\includegraphics[width=0.6\textwidth]{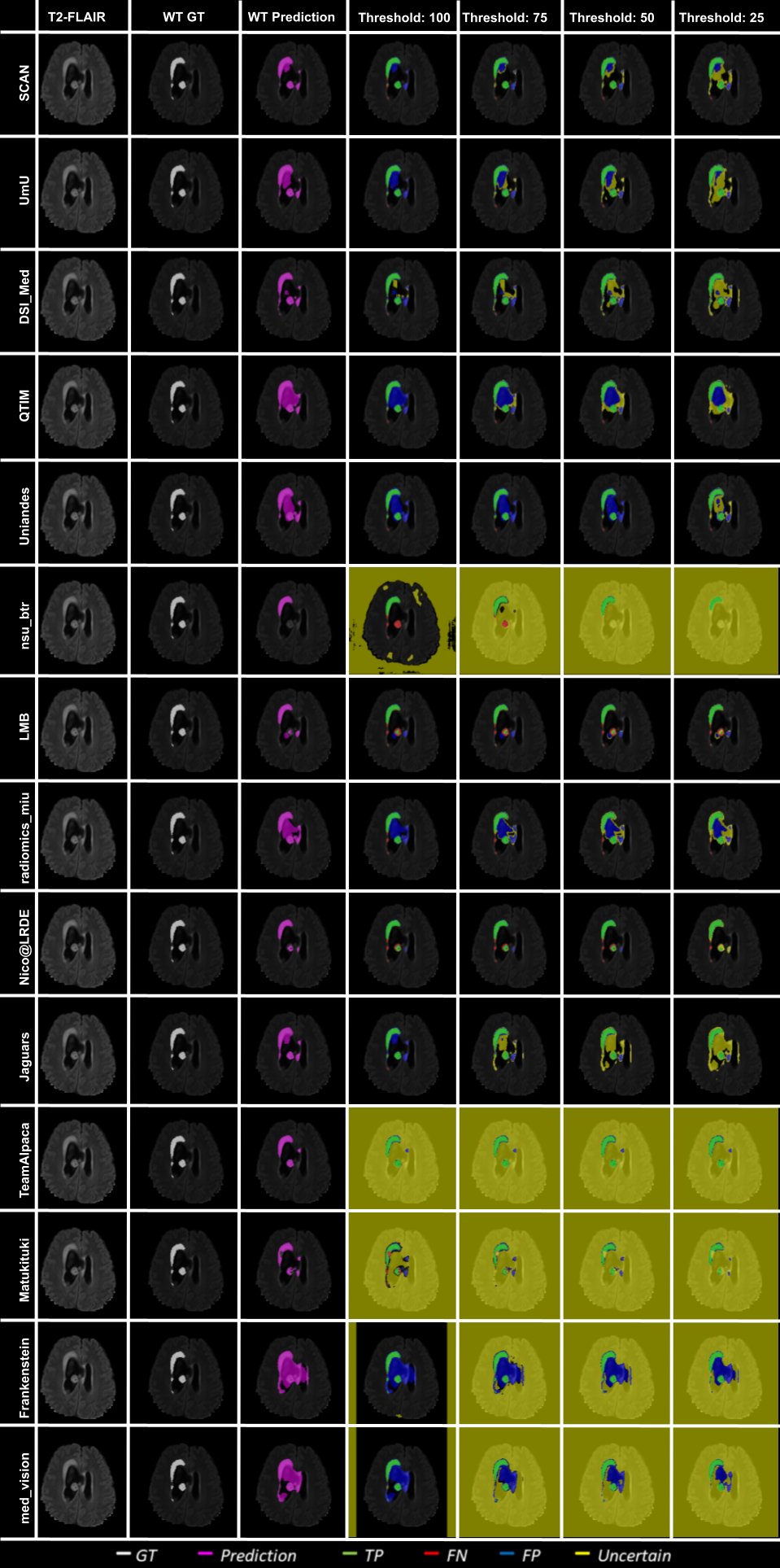}
\center \caption{Effect of uncertainty thresholding on a BraTS 2020 test case for whole tumor segmentation across different participating teams. (a) T2-FLAIR MRI (b) Ground Truth (c) Prediction (d) No filtering. Uncertainty Threshold = 100 (e) Uncertainty Threshold = 75 (f) Uncertainty Threshold = 50 (g) Uncertainty Threshold = 25.}
\label{fig:Qual_whole_017_098}
\end{figure*}

\begin{figure*}[t]
\centering
\includegraphics[width=0.6\textwidth]{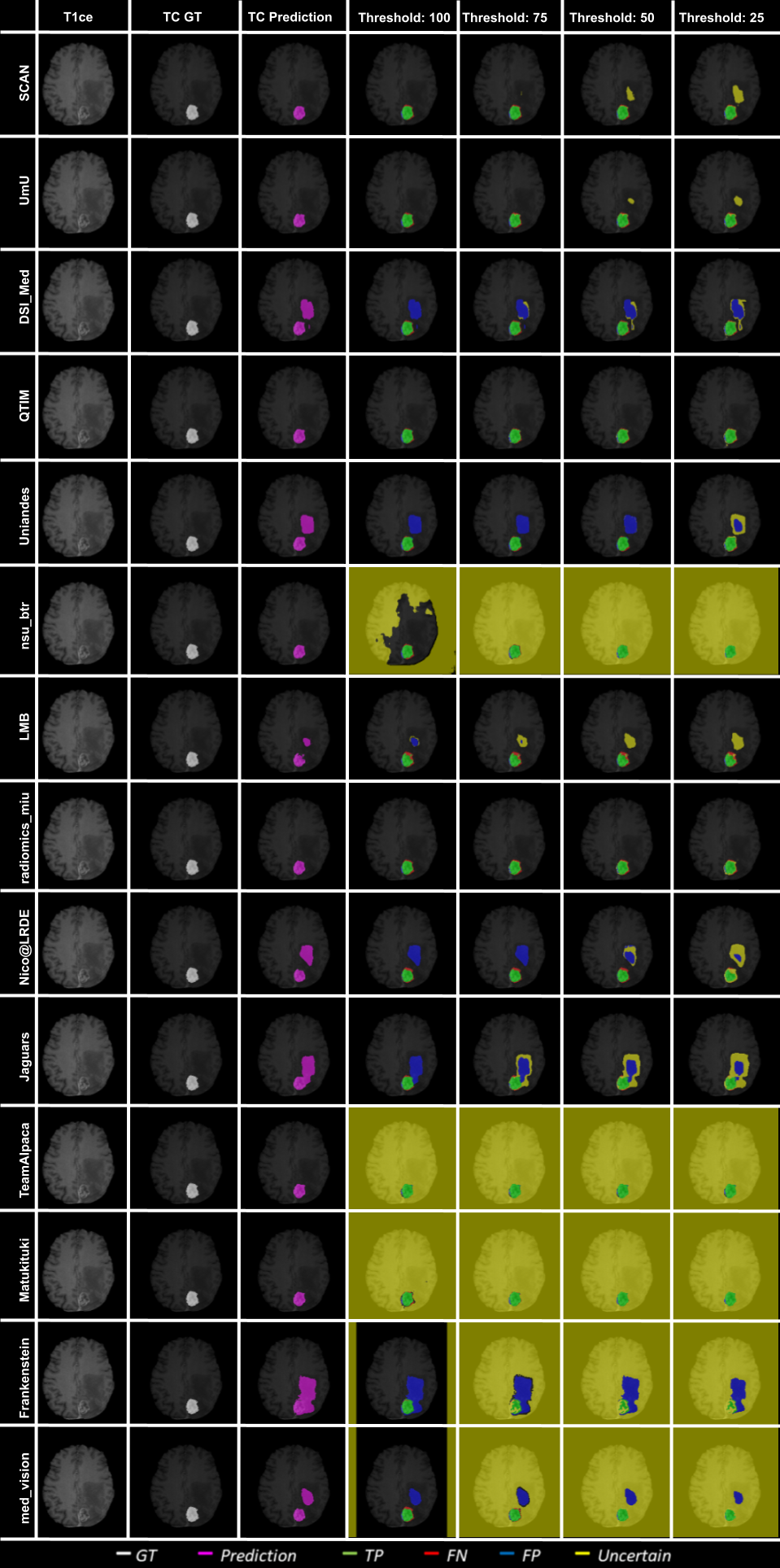} 
\center \caption{Effect of uncertainty thresholding on a BraTS 2020 test case for core tumor segmentation across different participating teams. (a) T1ce MRI (b) Ground Truth (c) Prediction (d) No filtering. Uncertainty Threshold = 100 (e) Uncertainty Threshold = 75 (f) Uncertainty Threshold = 50 (g) Uncertainty Threshold = 25.}
\label{fig:Qual_core_001_102}
\end{figure*}

\begin{figure*}[t]
\centering
\includegraphics[width=0.6\textwidth]{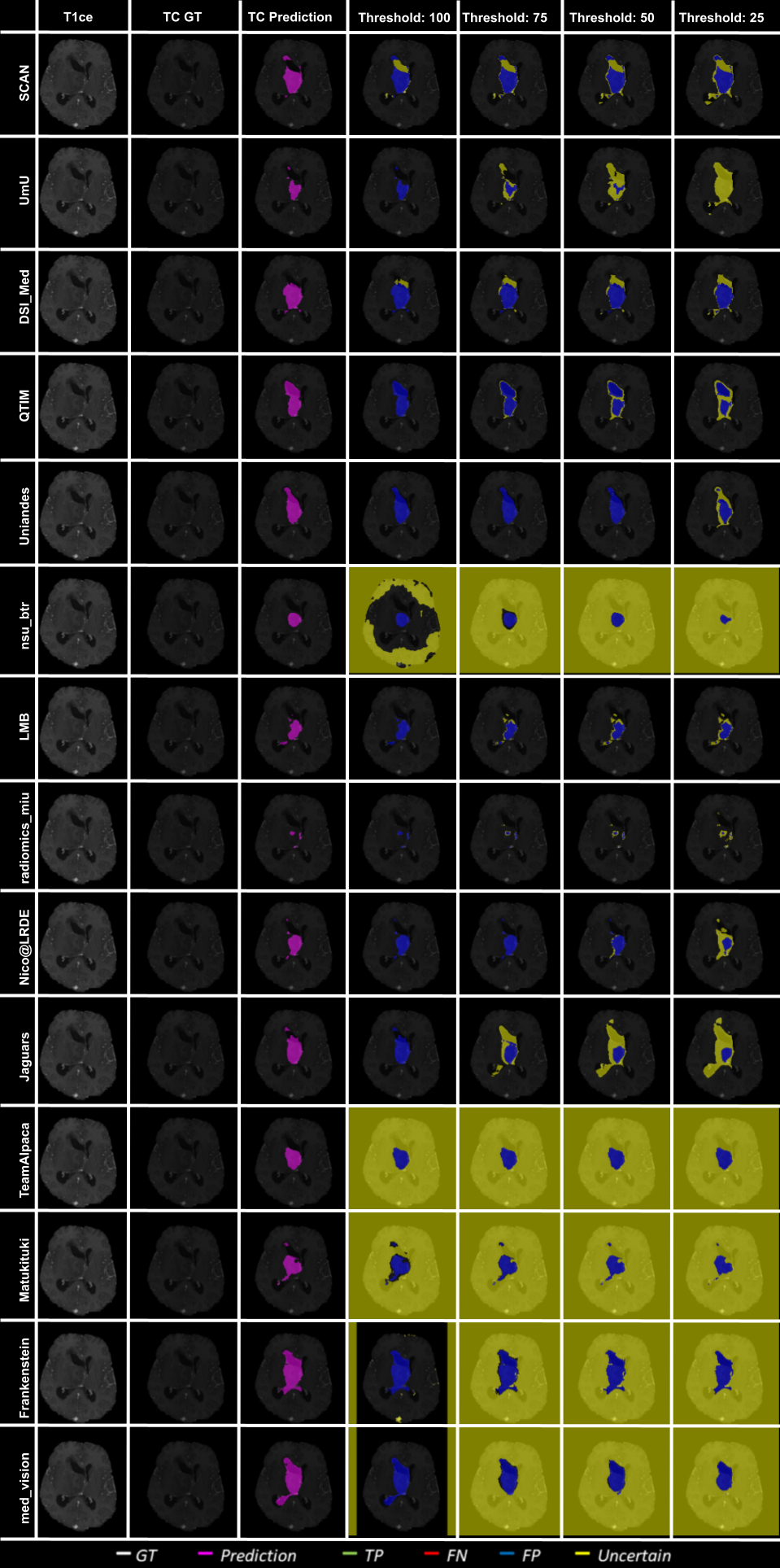} 
\center \caption{Effect of uncertainty thresholding on a BraTS 2020 test case for core tumor segmentation across different participating teams. (a) T1ce MRI (b) Ground Truth (c) Prediction (d) No filtering. Uncertainty Threshold = 100 (e) Uncertainty Threshold = 75 (f) Uncertainty Threshold = 50 (g) Uncertainty Threshold = 25.}
\label{fig:Qual_core_017_076}
\end{figure*}

\subsection{Qualitative Analysis} \label{QualAnalysis}
Figure~\ref{fig:Qual_whole_017_066} - Figure~\ref{fig:Qual_enhance_135_100} plots the effect of uncertainty threshold based filtering on example slices from a few BraTS 2020 test cases for all participating teams. Green voxels represent True Positive predictions, while blue and red voxels represent False Positive and False Negative predictions. We filter out voxels at different thresholds (100, 75, 50, and 25). Filtered voxels are marked as yellow. According to the developed uncertainty evaluation score (Section~\ref{UncMetric}), we want methods that filter out (marked as yellow) false positive and false negative voxels while retaining true positive and true negative voxels as we decrease the uncertainty threshold.\\

In Figure~\ref{fig:Qual_whole_017_066}, we visualize the effect of uncertainty based thresholding for WT segmentation on a single slice of a BraTS 2020 test case. A closer look at some of the better performing teams like \textit{Team SCAN}, \textit{Team UmU}, and \textit{Team DSI\_Med} reveals that these teams filter out more False Positives and False Negatives at a higher threshold than other teams like \textit{Team QTIM} and \textit{Team Uniandes}. We can also observe that lower-performing teams like \textit{Team Alpaca}, \textit{Team Matukituki}, \textit{Team Frankenstein}, and \textit{Team med\_vision} mark all background voxels as uncertain at a low threshold. As mentioned before, marking background voxels as uncertain is problematic as it shows that the method is not confident in its healthy-tissue segmentation and requires clinicians to review the segmentation. \\

In Figure~\ref{fig:Qual_whole_017_098}, we plot the effect of uncertainty based thresholding for WT segmentation on another slice of the same BraTS 2020 test case. Here we observe a similar trend where higher ranked teams can filter out False Positives and False Negatives at a higher threshold than other teams. \textit{Team SCAN} only filters negative predictions. This results in them never filtering out their False Positive predictions of the whole tumor inside the ventricles. It is problematic in a real-world scenario as we do not want a method that is over-confident about its positive pathology segmentation predictions. \\

Figure~\ref{fig:Qual_core_001_102} shows an example slice of a different BraTS 2020 patient and visualize the effect of uncertainty thresholding for core tumor segmentation. The figure highlights that team ranking is different across different cases as we can see that \textit{Team SCAN} and \textit{Team UmU} has similar prediction at \textit{Threshold:100}. However, \textit{Team SCAN} starts filtering out more true negatives sooner compared to \textit{Team UmU}, which would result in \textit{Team SCAN} ranked lower compared to \textit{Team UmU} for this particular BraTS test case. We can observe a similar trend when comparing \textit{Team DSI\_Med} and \textit{Team LMB}, where \textit{Team LMB} starts filtering out more false positives sooner than \textit{Team DSI\_Med}. Similarly, in Figure~\ref{fig:Qual_core_017_076}, we can observe that in scenarios where all teams are making errors by predicting a high amount of false positives, the overall uncertainty score would be more reliant on which teams can filter out these false positives sooner. For example, \textit{Team UmU} performs better compared to \textit{Team DSI\_Med}.\\

Figure~\ref{fig:Qual_enhance_135_100} depicts an example slice of uncertainty threshold based filtering for ET segmentation. Here we can see that when all teams make almost the same predictions with a high amount of true positives compared to false positives/false negatives, the overall uncertainty score is similar across teams. Except for teams that mark all background (healthy-tissue) voxels as uncertain, they perform poorly on the final score.

\section{Discussion}
This paper introduced a new score for evaluating uncertainties in the task of brain tumor segmentation during the BraTS 2020 challenge. The proposed score was used to rank different participating teams from the Uncertainty Quantification task of the BraTS 2020 challenge (QU-BraTS 2020).\\

The proposed evaluation score was developed with the clinical objective of enabling the clinician to review only the uncertain areas of an automatic segmentation algorithm instead of the complete segmentation. Toward this end, this score would reward algorithms that are confident when correct and uncertain when incorrect. The objective was evaluated by filtering (marking as uncertain) voxels with uncertainty higher than a specified threshold as uncertain. The task-dependent $DSC$ is measured only on the remaining unfiltered voxels. To ensure that method does not filter out a high number of correctly predicted voxels in order to achieve a better $DSC$, the developed evaluation score also keeps track of the number of filtered True Positive and True Negative voxels. Keeping track of these filtered TP and TN voxels ensures that the burden on the reviewing clinicians is not increased substantially. In short, the proposed score calculates the task-dependent metric score (i.e. $DSC$ for segmentation), the percentage of filtered true positives and true negatives at different uncertainty thresholds. It combines them to generate a single evaluation score for a single subject. \\

The analysis (Section~\ref{sec:teamrank}) of algorithms developed by the participating teams from the QU-BraTS 2020 task highlighted that the relative ranking of the participating teams for both the segmentation and uncertainty quantification tasks are different. The different ranking orders show that performing better on the segmentation task does not guarantee good performance on the uncertainty quantification task. An automatic segmentation method that provides both the segmentation and its uncertainties is more clinically relevant. Both the segmentation and the associated uncertainties provide complementary information. For example, automatic segmentation can provide accurate results with minimal clinician input. In contrast, the associated uncertainty would allow clinicians to see where to trust and review the segmentation before deploying it in clinical practice. \\

Results in Section~\ref{sec:tsrank} indicate that it is necessary to rank teams individually for each tumor entity as they rank differently across these entities. An ablation study on the proposed score (Section~\ref{sec:ablation}) showed the necessity of utilizing all three components ($DSC$, percentage of Filtered True Positive, and percentage of Filtered True Negative) for the proposed uncertainty evaluation score. \\

One of the significant limitations of the current analysis is the dependency between the segmentation results and the uncertainty generation methods, which does not allow for more in-depth analysis. It would be interesting to analyze and compare different uncertainty generation methods (e.g., \textit{Team SCAN}, \textit{Team UmU}, \textit{Team Alpaca}) when the segmentation method is the same across them. \\

We also observe a limitation of the proposed evaluation score. \textit{Team SCAN} performs better on the overall score by not marking any positive prediction as uncertain. In a real-world scenario, a method that is always confident about its positive predictions leads to confident over-segmentation. This shows that the developed uncertainty evaluation score is not perfect, and we need to keep improving it. We observed a similar trend in a recently conducted Probabilistic~Object~Detection challenge \citep{PODchallenge}, where the winning team attained the best score despite not using a probabilistic method. These two examples show the need to keep improving the developed task-depended uncertainty evaluation score for different tasks. \\

The $DSC$ is a good segmentation metric when the interest structure contains a high number of voxels. However, it is not a stable metric when calculated on a low number of voxels \citep{CommonLimit}. In the developed evaluation score, instability of the $DSC$ leads to low performance at a lower threshold (more filtered voxels), as $DSC$ calculation considers only a few remaining unfiltered voxels (Figure~\ref{fig:UncEntropy}). The poor stability of $DSC$ is a well-known challenge in the literature \citep{CommonLimit}. Future work could explore a task-dependent metric that is more stable across different uncertainty thresholds (i.e., different volumes for the structure of interest). For example, we can calculate Precision ($\frac{TP}{TP + FP}$) and Recall ($\frac{TP}{TP + FN}$) at different uncertainty thresholds and calculate the AUC of these curves (Precision vs Uncertainty threshold, and Recall vs Uncertainty threshold). A high-performing team should get a high AUC for both Precision and Recall (same as AUC for $DSC$). To achieve a high AUC for Precision, participating teams have to reduce FP (mark them as uncertain). Similarly, to attain a high AUC for Recall, participating teams have to reduce FN (mark them as uncertain). In this way, we can penalize teams that are highly confident in their positive predictions as well as those that are highly confident in their false negative predictions. \\

The proposed evaluation framework evaluates uncertainties for each tumor entity as a single class segmentation/uncertainty problem, while the overall tumor segmentation is a multi-class problem. Future extensions could involve developing methods to evaluate uncertainties in multi-class segmentation. Multi-class segmentation uncertainties and single-class segmentation uncertainties are different and can lead to different outcomes~\citep{CamarasaMELBA}. In addition,  the current evaluation framework focuses on filtering individual voxels, as most of the developed uncertainty frameworks generate per-voxel uncertainties that are not spatially correlated \citep{MCD, DeepEnsemble}. The recent development of spatially correlated uncertainty generation methods \citep{SSN} indicates the necessity of developing uncertainty evaluation scores that consider the spatial correlation between pixels/voxels. \\

Another future direction is obtaining "ground-truth" uncertainty maps and evaluating automatic uncertainty generation methods against these maps. One recent promising direction uses inter-observer and intra-rater variation to proxy for "ground-truth" uncertainty \citep{ProbUNet, PHiSeg, QUBIQ, VISCERAL}. One limitation of this approach is that it assumes that "ground-truth" uncertainties can be estimated through multiple labels provided by different raters for the same (often small) set of images. In recent papers \citep{zech2018variable, sheller2020federated}, it was noted that institutional biases \citep{mccarthy2016racial} play an essential factor in deep learning medical imaging model performance. However, variability in labeling across raters reflecting institutional biases are not direct proxies for "ground-truth" uncertainties. To expand on this point, inter-rater and intra-rater variability relies on the assumption of attaining a unique label. However, there are many situations where a unique label cannot necessarily be attained in some regions of an image. For example, at boundaries between tumor and healthy tissue in MRI due partly to partial volume effects but also because the labels cannot be seen in the MRI (and cannot be verified without a biopsy in the case of a tumour). For the latter case, each annotator is "guessing" the location of the boundary when none are confident in their annotations. The result might be measuring contextual rater biases (e.g., based on their radiology backgrounds) but not reflecting the true uncertainties in the labels themselves (e.g., whether a particular pixel is an enhancing tumour). One alternative approach could be asking annotators to mark areas they are not certain about, such as tumor boundaries in an MRI scan. These "uncertain" areas can then serve as "ground-truth," and uncertainty estimates generated by algorithms can be compared to it. That being said, acquiring a "ground-truth" uncertainty is still an open area of research.\\

The approach developed for QU-BraTS has shown promising results in different applications. For example, the proposed score was used to evaluate uncertainties for brain tumor segmentation when a crucial MR sequence is missing~\citep{HADNet}. The proposed score has also been used to evaluate multi-class segmentation of the carotid artery lumen and the vessel wall \citep{CamarasaMELBA}.

%%%%%%%%%%%%%%%%%%%%%%%%%%%%%%%%%%%%%%%%%%%%%%%%%%%%%%%%%%%%%%%%%%%%%%%
% Mandatory Sections. Please complete, especially for final publication
%%%%%%%%%%%%%%%%%%%%%%%%%%%%%%%%%%%%%%%%%%%%%%%%%%%%%%%%%%%%%%%%%%%%%%%

% Acknowledgements. 
% Please include any funding, intellectual contributions not included in the authorship, and any other acknowledgements. 
\acks{Research reported in this publication was partly supported by the Informatics Technology for Cancer Research (ITCR) program of the National Cancer Institute (NCI) of the National Institutes of Health (NIH), under award numbers NIH/NCI/ITCR:U01CA242871 and NIH/NCI/ITCR:U24CA189523. It was also partly supported by the National Institute of Neurological Disorders and Stroke (NINDS) of the NIH, under award number NIH/NINDS:R01NS042645, The content of this publication is solely the responsibility of the authors and does not represent the official views of the NIH. \\

This work was supported by a Canadian Natural Science and Engineering Research Council
(NSERC) Collaborative Research and Development Grant (CRDPJ 505357 - 16), Synaptive
Medical, and the Canada Institute for Advanced Research (CIFAR) Artificial Intelligence
Chairs program.}

% Ethical Standards.
% Please edit with the appropriate ethics considerations for your work. Include any pertinent IRB information, etc.
% 
% Please note that the submission requirements included:
% The work presented must follow appropriate ethical standards in conducting research and writing the manuscript, following all applicable laws and regulations regarding treatment of animals or human subjects.
\ethics{The work follows appropriate ethical standards in conducting research and writing the manuscript, following all applicable laws and regulations regarding treatment of animals or human subjects.}

% Conflict of Interest
% Declaration of possible conflicts of interest: Authors must disclose any financial, organisational, commercial or personal conflicts of interest that might bias their work.  
% If no conflicts, please say "We declare we don't have conflicts of interest."
\coi{The conflicts of interest have not been entered yet.}

% \newpage

% \vskip 0.2in
\bibliography{sample}

\newpage

\appendix % optional
\section*{Appendix A - Box plots for individual scores}

\begin{figure*}[t]
\centering
\includegraphics[width=0.45\textwidth]{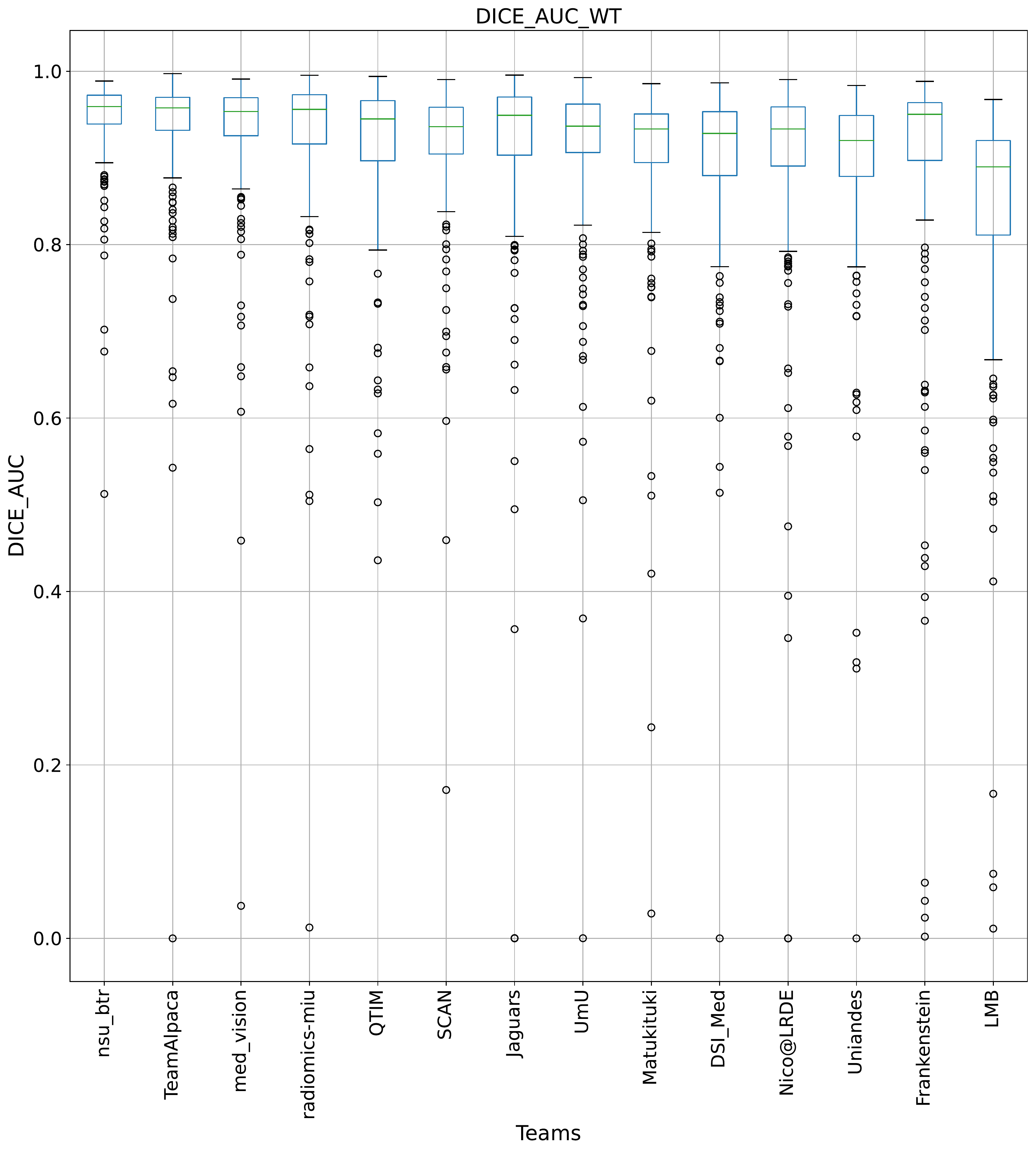} 
\center \caption{QU-BraTS 2020 boxplots depicting DICE\_AUC distribution for all teams across different participants for Whole Tumor on the BraTS 2020 test set (higher is better).}
\label{fig:BP_DICE_AUC_WT}
\end{figure*}

\begin{figure*}[t]
\centering
\includegraphics[width=0.45\textwidth]{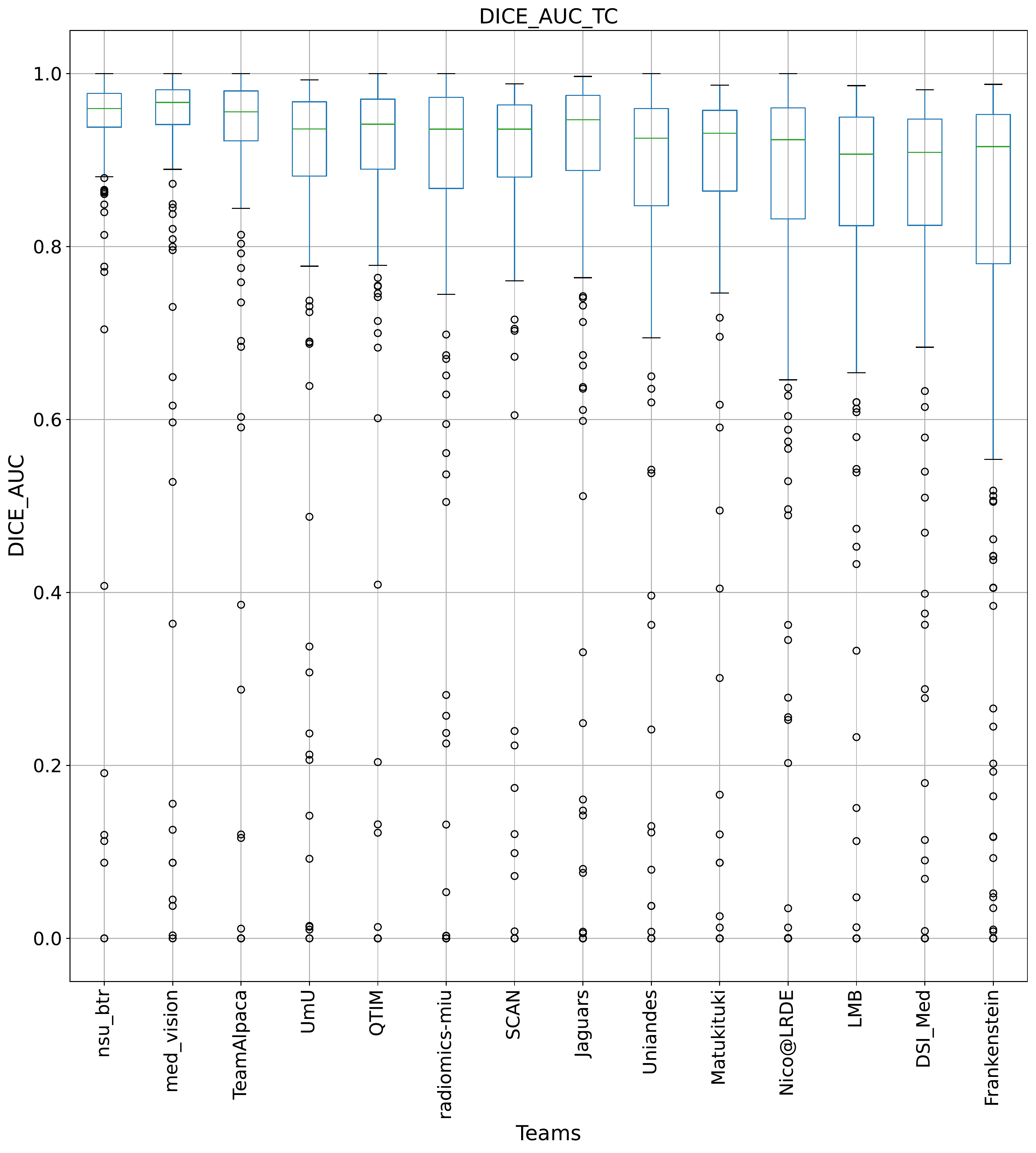} 
\center \caption{QU-BraTS 2020 boxplots depicting DICE\_AUC distribution for all teams across different participants for Tumor Core on the BraTS 2020 test set (higher is better).}
\label{fig:BP_DICE_AUC_TC}
\end{figure*}

\begin{figure*}[t]
\centering
\includegraphics[width=0.45\textwidth]{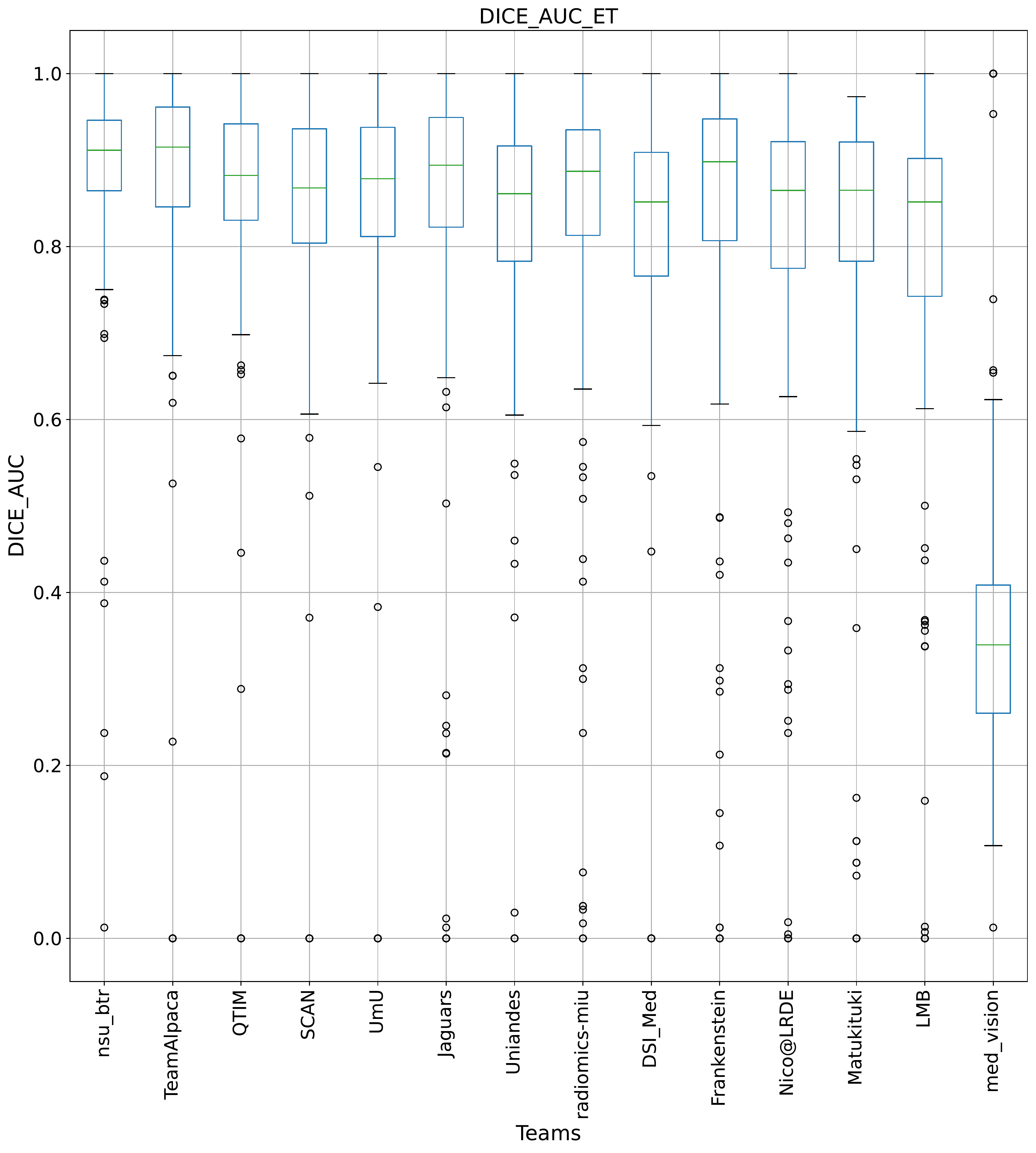} 
\center \caption{QU-BraTS 2020 boxplots depicting DICE\_AUC distribution  for all teams across different participants for Enhancing Tumor on the BraTS 2020 test set (higher is better).}
\label{fig:BP_DICE_AUC_ET}
\end{figure*}

\begin{figure*}[t]
\centering
\includegraphics[width=0.45\textwidth]{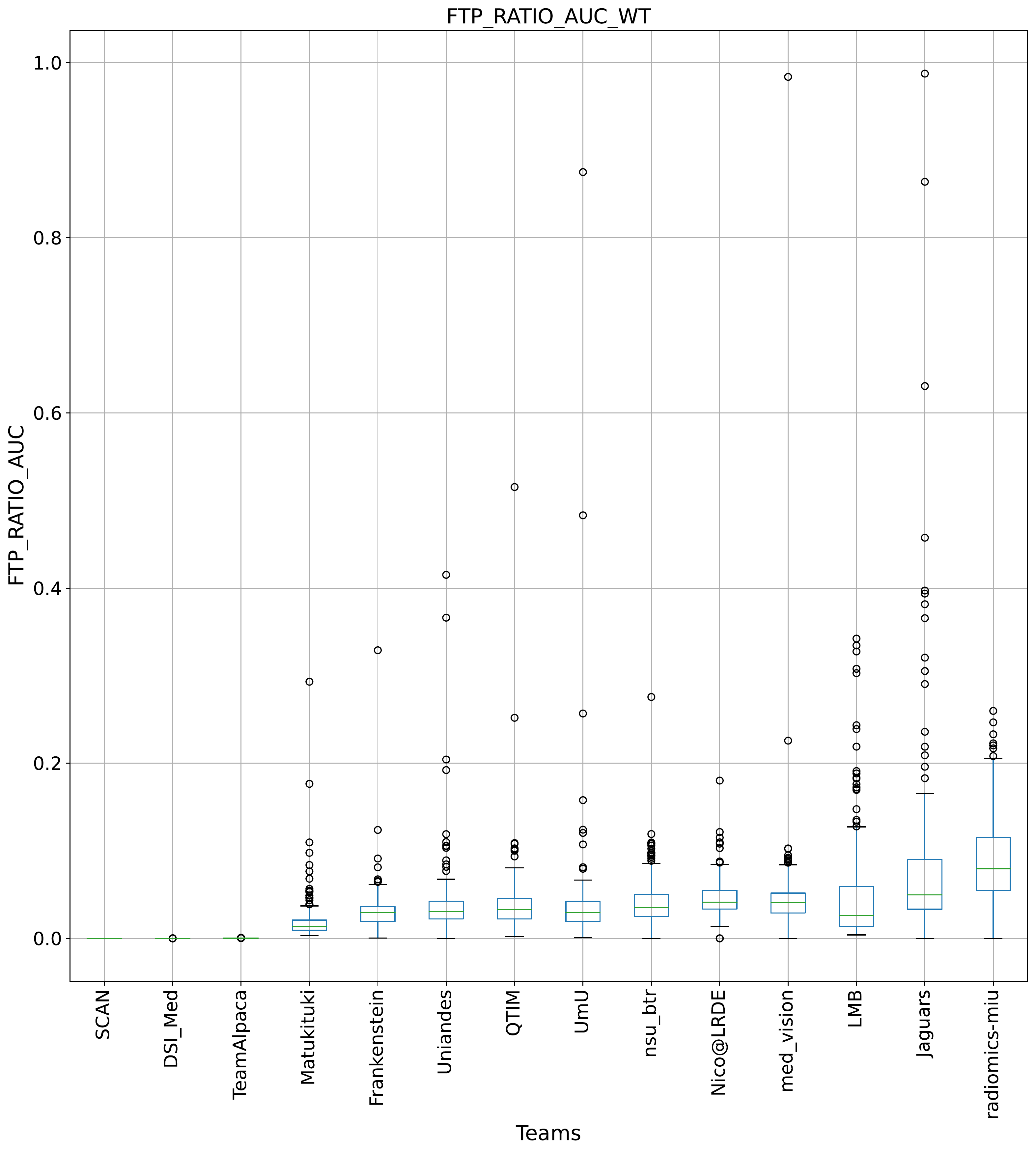} 
\center \caption{QU-BraTS 2020 boxplots depicting FTP\_RATIO\_AUC distribution for all teams across different participants for Whole Tumor on the BraTS 2020 test set (lower is better).}
\label{fig:BP_FTP_RATIO_AUC_WT}
\end{figure*}

\begin{figure*}[t]
\centering
\includegraphics[width=0.45\textwidth]{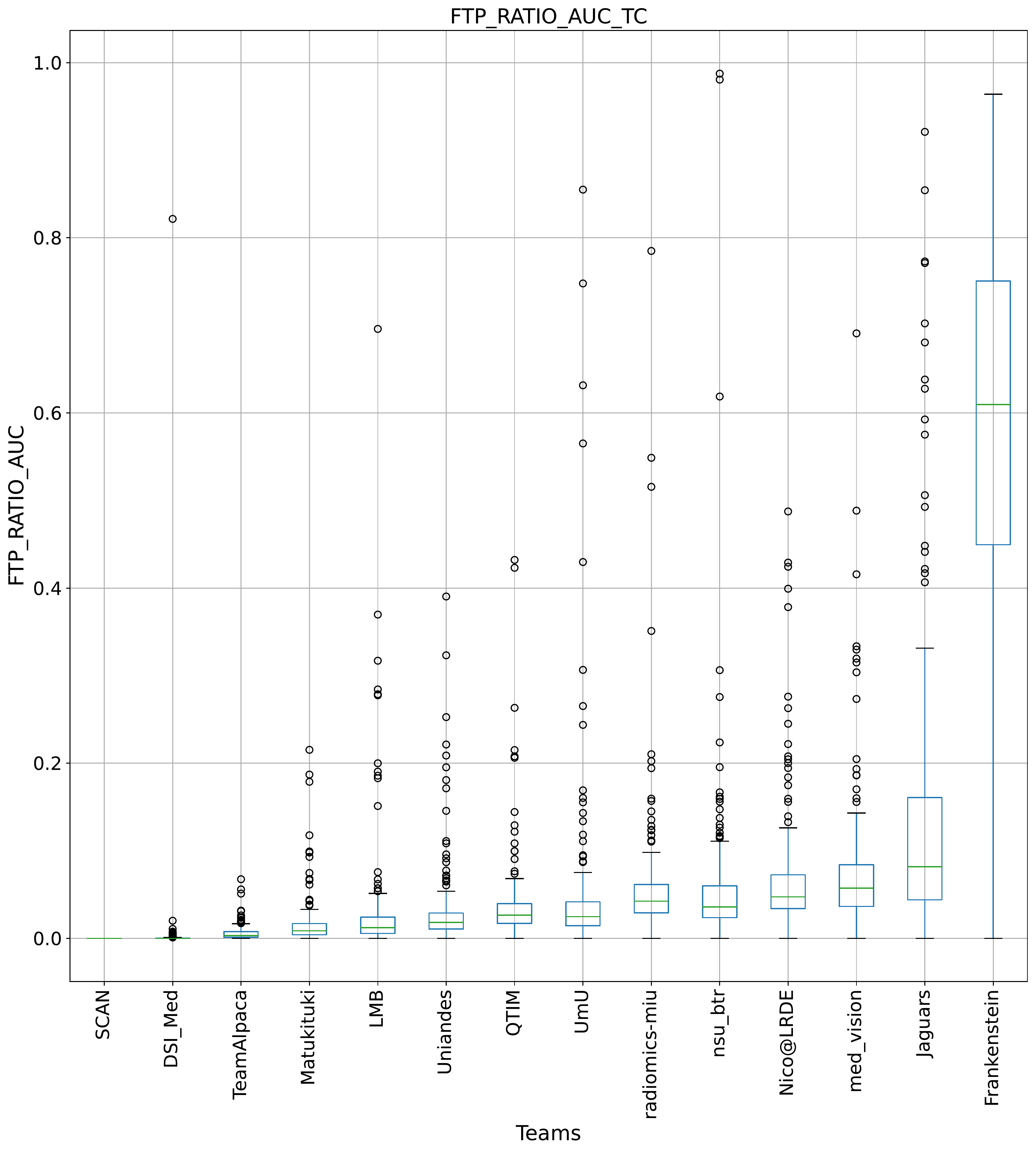} 
\center \caption{QU-BraTS 2020 boxplots depicting FTP\_RATIO\_AUC distribution for all teams across different participants for Tumor Core on the BraTS 2020 test set (lower is better).}
\label{fig:BP_FTP_RATIO_AUC_TC}
\end{figure*}

\begin{figure*}[t]
\centering
\includegraphics[width=0.45\textwidth]{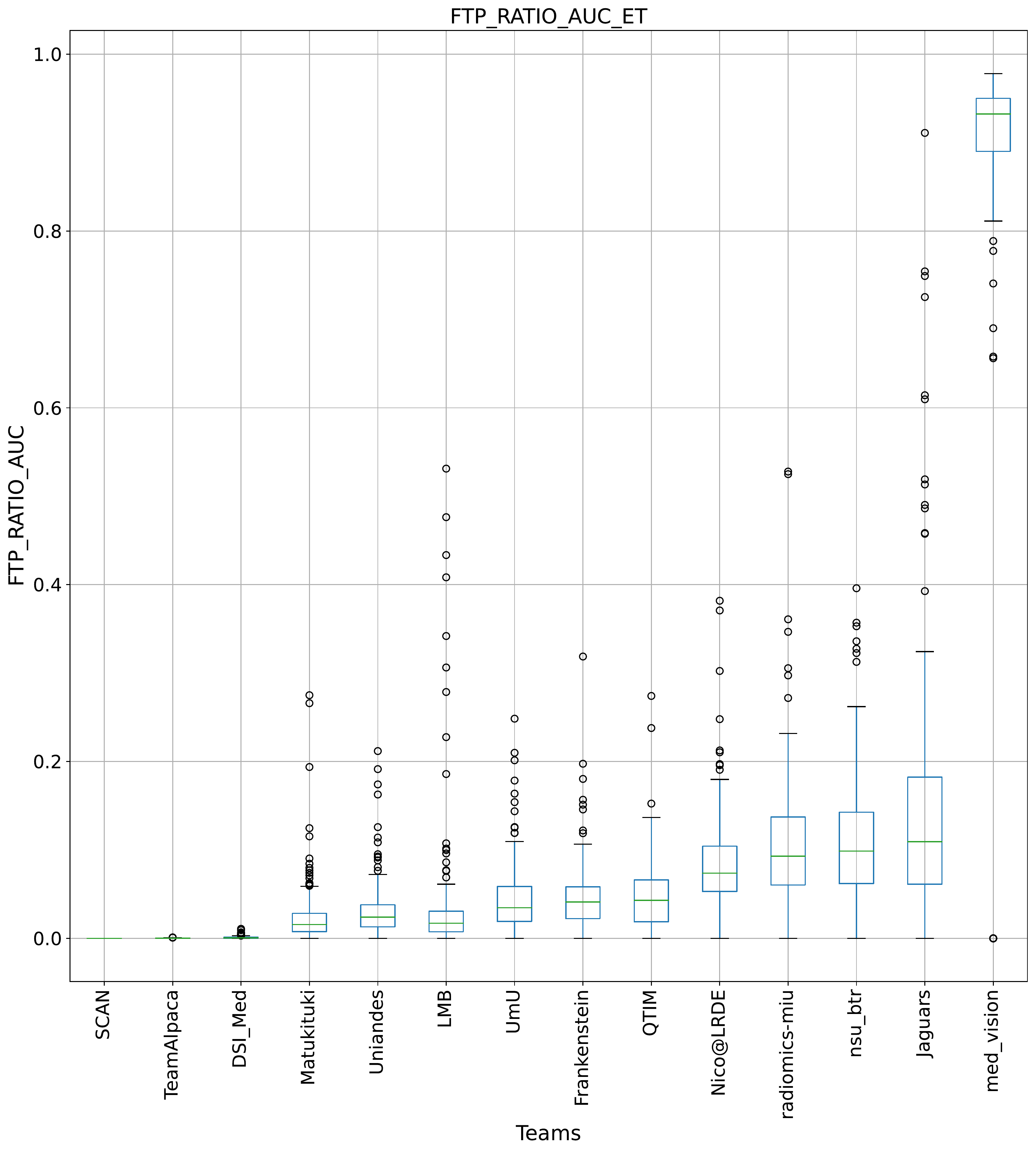} 
\center \caption{QU-BraTS 2020 boxplots depicting FTP\_RATIO\_AUC distribution for all teams across different participants for Enhancing Tumor on the BraTS 2020 test set (lower is better).}
\label{fig:BP_FTP_RATIO_AUC_ET}
\end{figure*}

\begin{figure*}[t]
\centering
\includegraphics[width=0.45\textwidth]{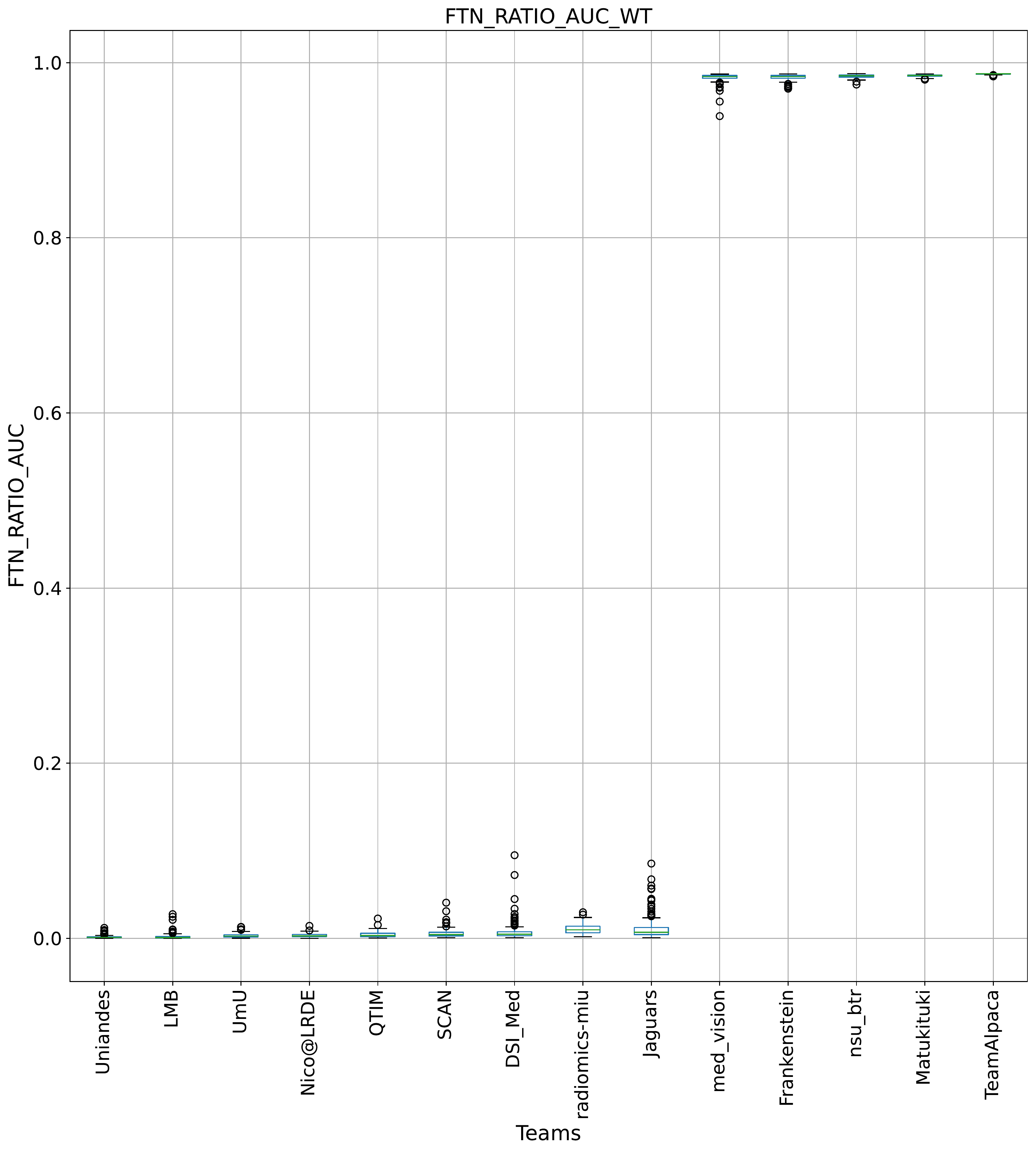} 
\center \caption{QU-BraTS 2020 boxplots depicting FTN\_RATIO\_AUC distribution for all teams across different participants for Whole Tumor on the BraTS 2020 test set (lower is better).}
\label{fig:BP_FTN_RATIO_AUC_WT}
\end{figure*}

\begin{figure*}[t]
\centering
\includegraphics[width=0.45\textwidth]{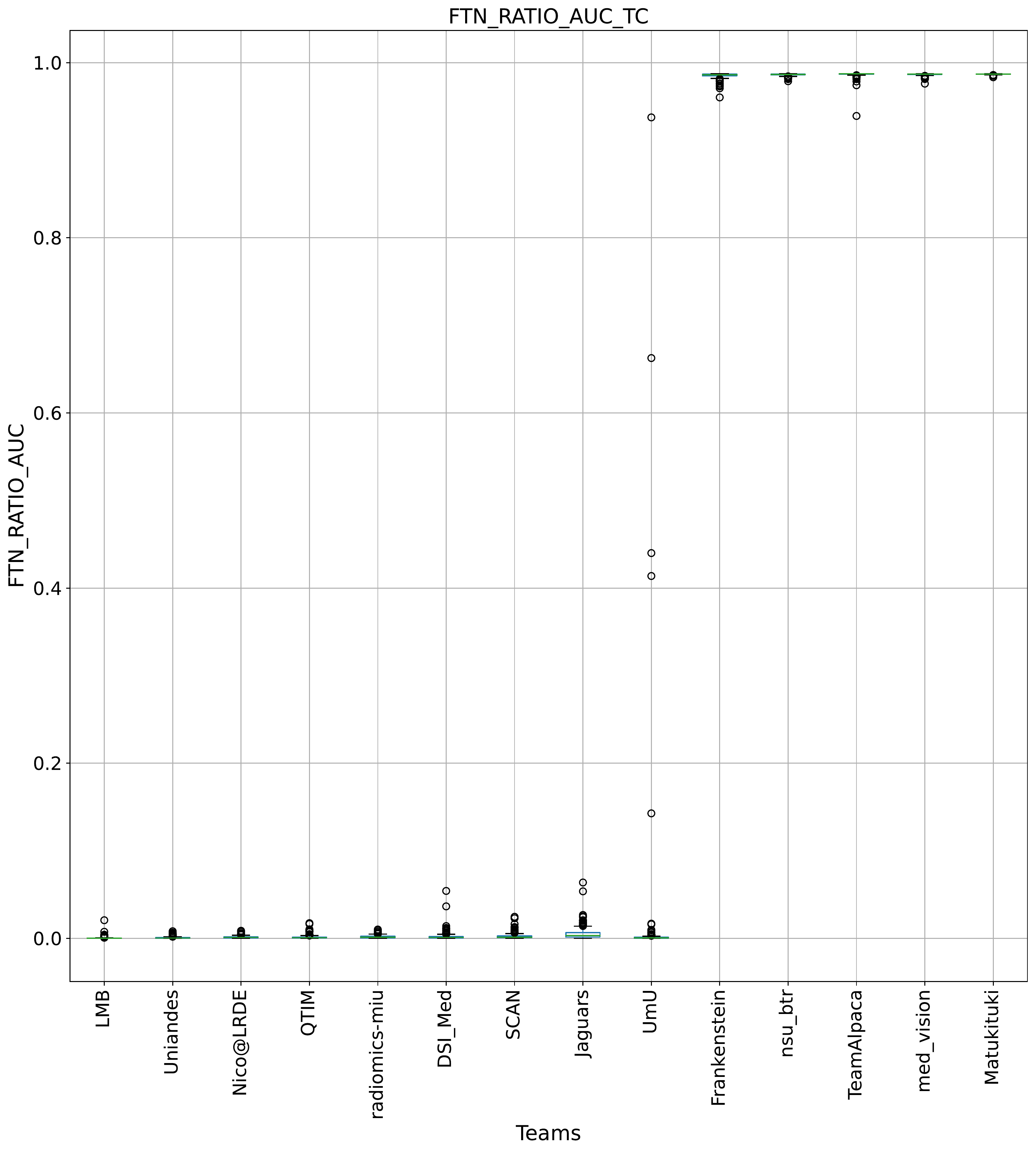} 
\center \caption{QU-BraTS 2020 boxplots depicting FTN\_RATIO\_AUC distribution for all teams across different participants for Tumor Core on the BraTS 2020 test set (lower is better).}
\label{fig:BP_FTN_RATIO_AUC_TC}
\end{figure*}

\begin{figure*}[t]
\centering
\includegraphics[width=0.45\textwidth]{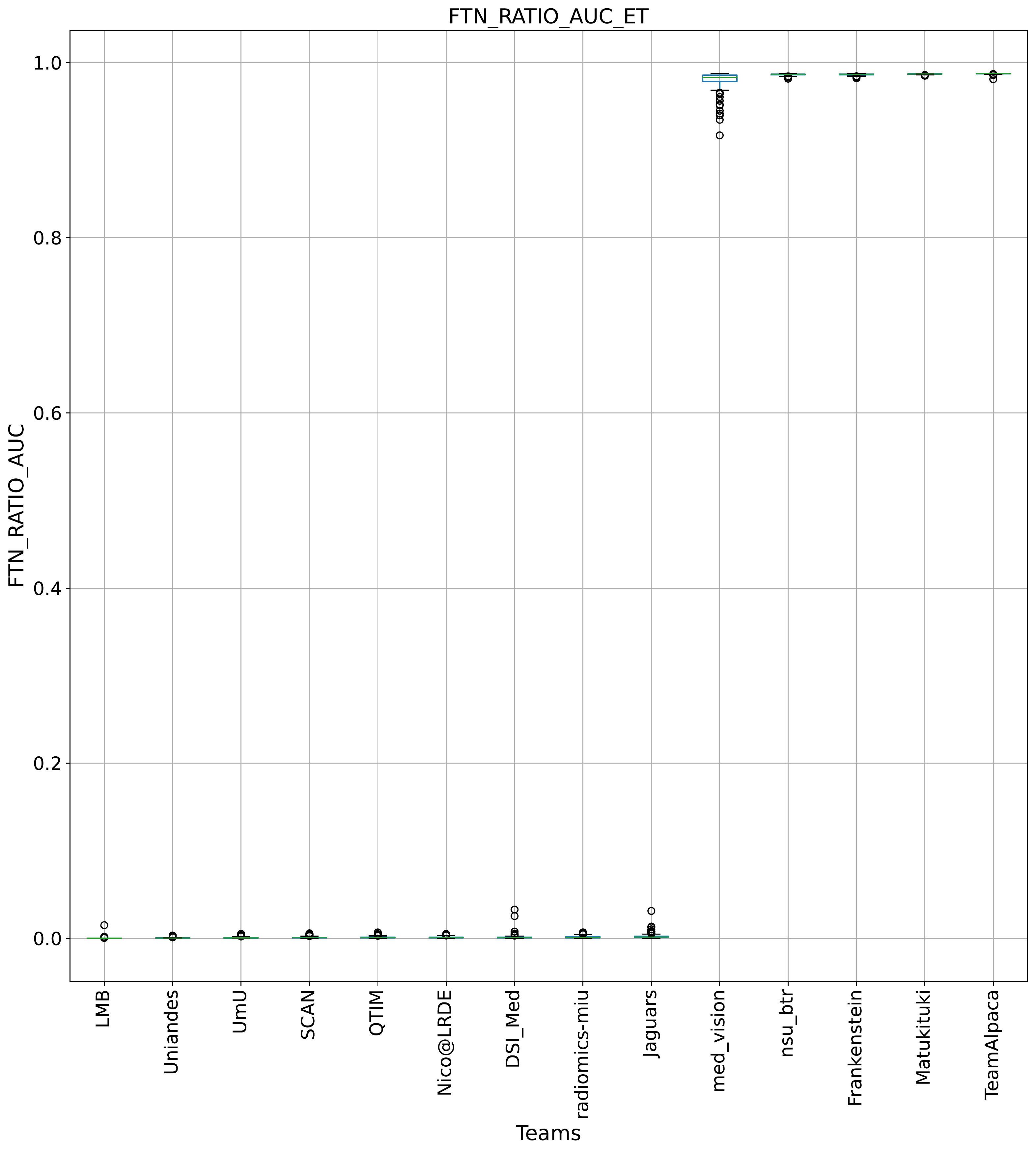} 
\center \caption{QU-BraTS 2020 boxplots depicting FTN\_RATIO\_AUC distribution for all teams across different participants for Enhancing Tumor on the BraTS 2020 test set (lower is better).}
\label{fig:BP_FTN_RATIO_AUC_ET}
\end{figure*}

\begin{figure*}[t]
\centering
\includegraphics[width=0.45\textwidth]{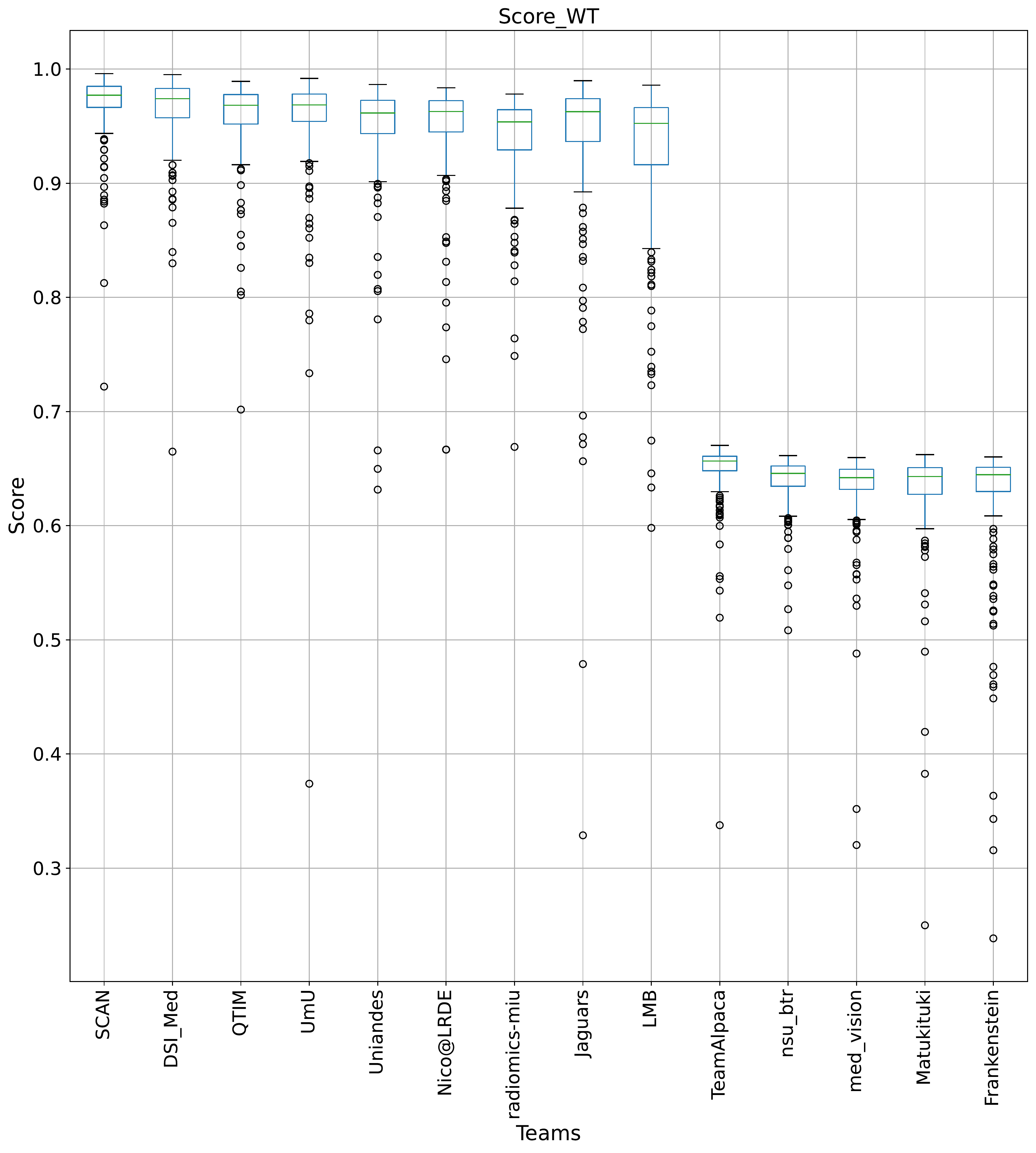}
\center \caption{QU-BraTS 2020 boxplots depicting Score distribution for all teams across different participants for Whole Tumor on the BraTS 2020 test set (higher is better).}
\label{fig:BP_Score_WT}
\end{figure*}

\begin{figure*}[t]
\centering
\includegraphics[width=0.45\textwidth]{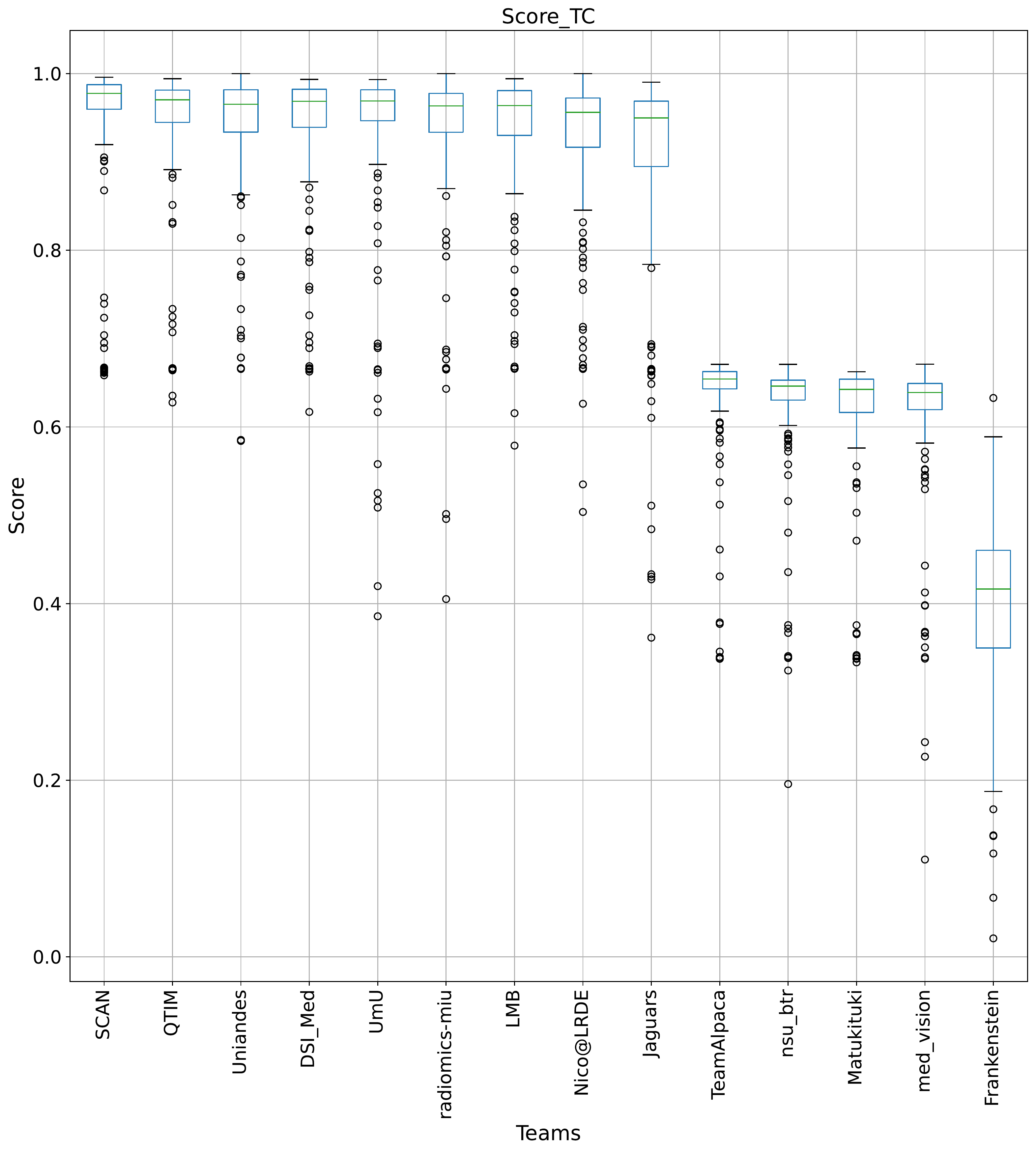} 
\center \caption{QU-BraTS 2020 boxplots depicting Score distribution for all teams across different participants for Tumor Core on the BraTS 2020 test set (higher is better).}
\label{fig:BP_Score_TC}
\end{figure*}

\begin{figure*}[t]
\centering
\includegraphics[width=0.45\textwidth]{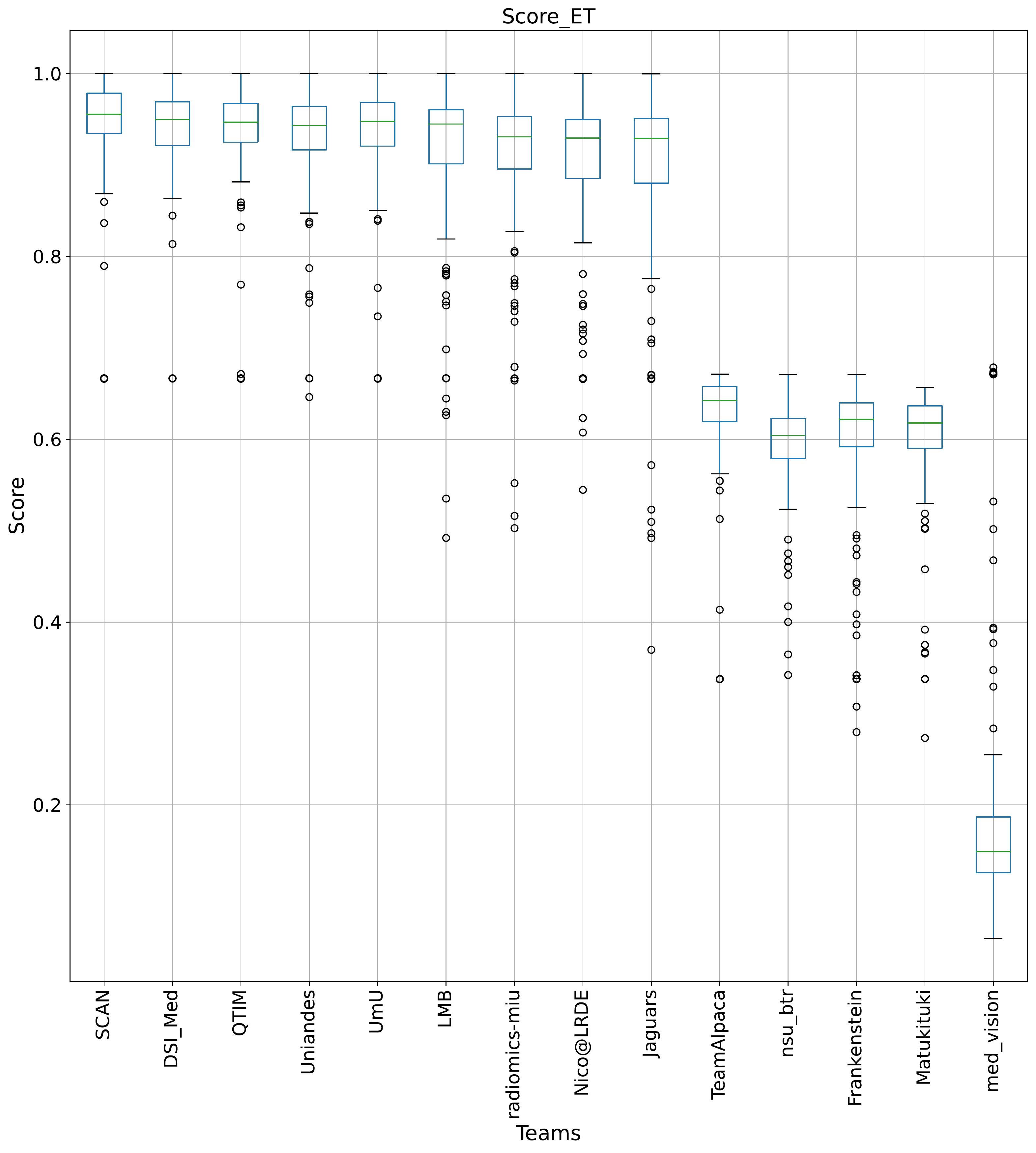} 
\center \caption{QU-BraTS 2020 boxplots depicting Score distribution for all teams across different participants for Enhancing Tumor on the BraTS 2020 test set (higher is better).}
\label{fig:BP_Score_ET}
\end{figure*}

This appendix provides box plots for four different scores (DICE\_AUC, FTP\_RATIO\_AUC, FTN\_RATIO\_AUC, and Score - Equation~\ref{Equation-score}) for three different tumor entities (WT, TC, and ET) for each team. The teams are ranked from better to worse performance according to mean values across all patients for each score. Higher is better for DICE\_AUC (Figure~\ref{fig:BP_DICE_AUC_WT} - Figure~\ref{fig:BP_DICE_AUC_ET}) and Score (Figure~\ref{fig:BP_Score_WT} - Figure~\ref{fig:BP_Score_ET}), while lower is better for FTP\_RATIO\_AUC (Figure~\ref{fig:BP_FTP_RATIO_AUC_WT} - Figure~\ref{fig:BP_FTP_RATIO_AUC_ET})  and FTN\_RATIO\_AUC (Figure~\ref{fig:BP_FTN_RATIO_AUC_WT} - Figure~\ref{fig:BP_FTN_RATIO_AUC_ET}). \\

Note that these box plots are different from ranking plots, as the ranking plots describe the overall performance across different tumor entities and different subjects as described in Section~\ref{RankScheme}. From these plots, we can see that while for all three tumor entity DICE\_AUC plots, \textit{Team nsu\_btr} performs better than other teams, their overall Score is comparatively lower than other teams as they do not perform well for FTP\_RATIO\_AUC and FTN\_RATIO\_AUC. \\

Similarly, we also observe that \textit{Team SCAN} does not outperform other teams for DICE\_AUC but comfortably outperforms other teams in FTP\_RATIO\_AUC. They perform relatively similar to other top-ranked teams in the FTN\_RATIO\_AUC score. Overall, they achieve the best performance for the Score across all three tumor entities. The main reason for them outperforming other teams for FTP\_RATIO\_AUC is how they developed their uncertainty generation method. They found that they achieved the best results on the given Score (Equation~\ref{Equation-score}) by considering all positive predictions as certain  (Section~\ref{TeamSCAN}). \\

In terms of overall Scores, we observe that \textit{Team SCAN} comfortably outperforms all other teams for each tumor entity. \textit{Team QTIM} and \textit{Team Uniandes} report better mean scores across different patients compared to \textit{Team SCAN}. Despite this, they do not achieve an overall better ranking for each patient, which shows the usefulness of reporting ranking and statistical-significance analysis across different patients rather than just reporting mean overall Score across patients.

\begin{figure*}[t]
\centering
\includegraphics[width=0.45\textwidth]{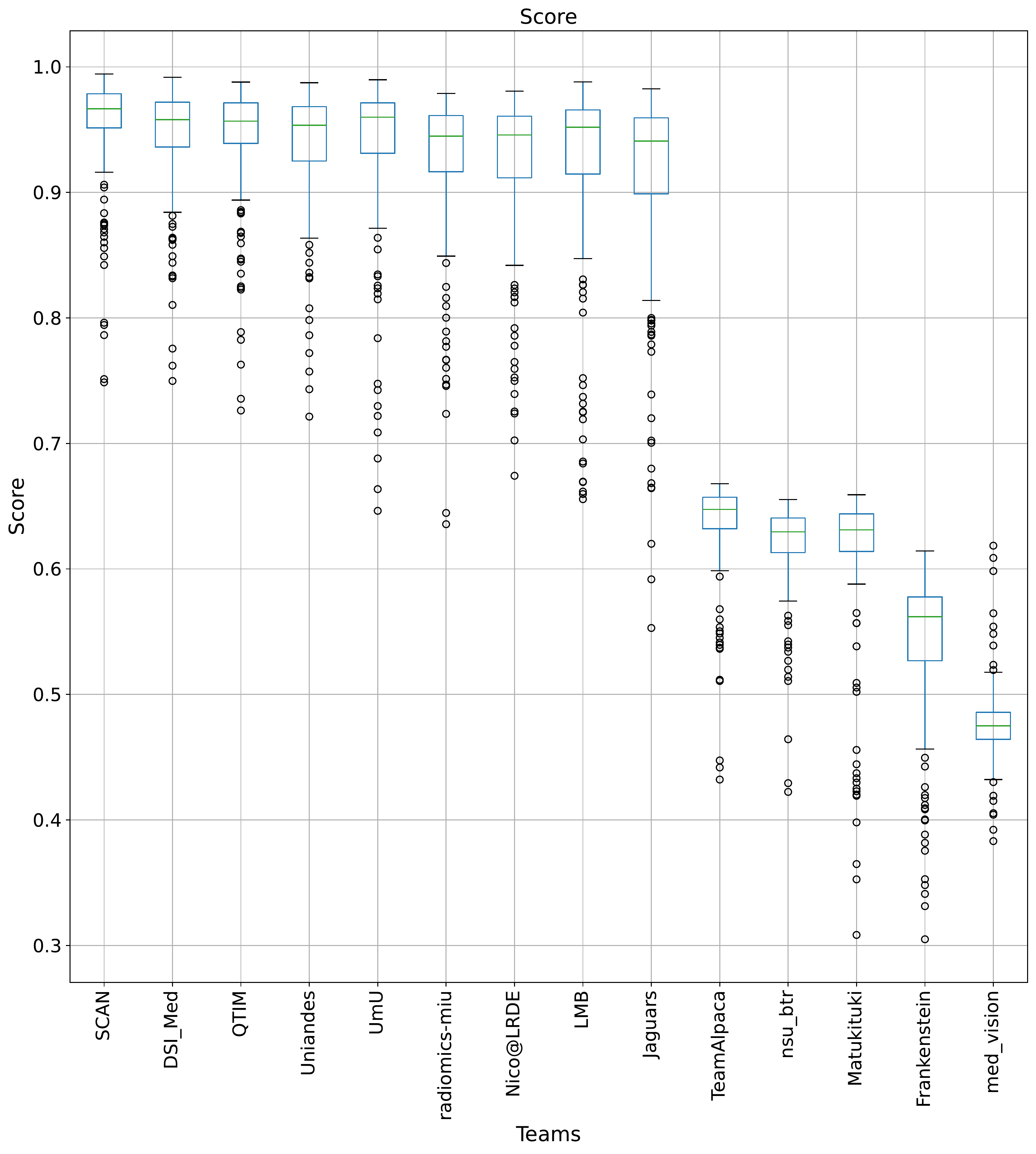} 
\center \caption{QU-BraTS 2020 boxplots depicting overall Score distribution for all teams across different participants on the BraTS 2020 test set (higher is better).}
\label{fig:BP_Score}
\end{figure*}

%%%%%%%%%%%%%%%%%%%%%%%%%%%%%%%%%%%%%%%%%%%%%%%%%%%%%%%%%%%%%%%%%%%%%%%%
%%%%%%
%%%%%%%%%%%%%%%%%%%%%%%%%%%%%%%%%%%%%%%%%%%%%%%%%%%%%%%%%%%%%%%%%%%%%%%
\appendix
\section*{Appendix B - QU-BraTS 2019} \label{AppenB}

In this appendix, we analyze and briefly describe methods employed by participating teams in BraTS 2019 sub-challenge on uncertainty quantification. A total of 15 teams participated in the challenge. From these 15 teams, five teams further participated during the following QU-BraTS 2020 challenge.  \\ 

\noindent \textbf{BraTS 2019 dataset:} As described in Section~\ref{3DUNetExp}, BraTS 2019 dataset contains 335 patient MRIs in the training set, 125 in the validation set, and 166 in the testing set. All teams developed their method using the training set and the validation set. Ground truth segmentation for the validation set was not publicly available for the teams. The final performance of all teams was measured on the testing set, where each team had access to a 48-hour window to upload their result to the server (\url{https://ipp.cbica.upenn.edu/}). \\

\noindent \textbf{QU-BraTS 2019 results on the test set:}  We ran the task of uncertainty quantification preliminary during the challenge and did not employ any ranking scheme. Also, the score used during the challenge was different from the one described in Section~\ref{UncMetric}. Precisely, we did not calculate the AUC of Ratio of Filtered True Negatives vs. Uncertainty threshold until the validation phase was ended; and only used AUCs of $DSC$ vs. Uncertainty Threshold and Ratio of Filtered True Positives vs. Uncertainty Threshold. After the validation phase, using qualitative inspection, we found that many teams were employing 1 - softmax confidence as an uncertainty measure, which is not helpful from a real clinical point of view as described in Section~\ref{UncMetric} and Section~\ref{QualAnalysis}. Keeping this in mind, we added the AUC of Ratio of Filtered True Negatives vs. Uncertainty threshold during the final testing phase. Table~\ref{tab:QUBraTS2019} lists all team names and their performance on the BraTS 2019 test phase. The table shows that teams that employed 1 - softmax\_confidence as uncertainty measure performed poorly on FTN\_RATIO\_AUC score (Ex. \textit{Team Alpaca}, \textit{Team DRAG}, \textit{Team ODU\_vision\_lab}, etc.). We want to point out that we did not employ the ranking strategy used in the QU-BraTS 2020 challenge during the QU-BraTS 2019 challenge. As we discussed in Appendix A, the ranking strategy and statistical significance analysis reflect the true potential of the method compared to just ranking teams according to their mean performance across testing cases.

\begin{landscape}
\begin{table}[]
\centering
\caption{Final performance on the BraTS 2019 testing dataset for teams participating in the preliminary challenge on quantification of uncertainty in brain tumor segmentation task. Here, mean values for each score across all patient in the testing dataset is listed.}
\resizebox{1.4\textwidth}{!}{%
\begin{tabular}{l|c|rrr|rrr|rrr|rrr|r}
\hline
\multirow{2}{*}{\textbf{Team}} & \multicolumn{1}{l|}{\multirow{2}{*}{\textbf{\#cases}}} & \multicolumn{3}{c|}{\textbf{DICE\_AUC}} & \multicolumn{3}{c|}{\textbf{FTP\_RATIO\_AUC}} & \multicolumn{3}{c|}{\textbf{FTN\_RATIO\_AUC}} & \multicolumn{3}{c|}{\textbf{Score}}     & \multicolumn{1}{l}{\multirow{2}{*}{\textbf{\begin{tabular}[c]{@{}l@{}}overall\\ Score\end{tabular}}}} \\ \cline{3-14}
                      & \multicolumn{1}{l|}{}                         & \textbf{WT} & \textbf{TC} & \textbf{ET} & \textbf{WT}   & \textbf{TC}   & \textbf{ET}   & \textbf{WT}   & \textbf{TC}   & \textbf{ET}   & \textbf{WT} & \textbf{TC} & \textbf{ET} & \multicolumn{1}{l}{}                                                                                  \\ \hline
\textbf{SCAN \citep{SCAN2020}}                  & 166                                           & 0.8837      & 0.8253      & 0.8209      & 0.0358        & 0.0771        & 0.14958       & 0.01919       & 0.0076        & 0.0060        & 0.9429      & 0.9135      & 0.8885      & 0.9150                                                                                                \\
\textbf{RADIOMICS-MIU \citep{RadiomicsMIU}}         & 166                                           & 0.8595      & 0.8122      & 0.7759      & 0.0421        & 0.0906        & 0.12009       & 0.00380       & 0.0012        & 0.0008        & 0.9379      & 0.9068      & 0.8850      & 0.9099                                                                                                \\
\textbf{UmU \citep{UmU2019}}                   & 166                                           & 0.8520      & 0.8077      & 0.7892      & 0.0602        & 0.1229        & 0.14089       & 0.00334       & 0.0150        & 0.0010        & 0.9295      & 0.8899      & 0.8824      & 0.9006                                                                                                \\
\textbf{xuefeng \citep{xuefeng2019}}               & 166                                           & 0.8746      & 0.8432      & 0.8120      & 0.0894        & 0.1642        & 0.27216       & 0.00969       & 0.0049        & 0.0024        & 0.9252      & 0.8914      & 0.8458      & 0.8874                                                                                                \\
\textbf{UTintelligence \citep{UTintelligence2019}}        & 162                                           & 0.7800      & 0.6787      & 0.6688      & 0.0117        & 0.0528        & 0.12901       & 0.00000       & 0.0000        & 0.0000        & 0.9228      & 0.8753      & 0.8466      & 0.8816                                                                                                \\
\textbf{NVDLMED \citep{NVDLMED2019}}               & 166                                           & 0.8651      & 0.8203      & 0.8251      & 0.0213        & 0.0679        & 0.10958       & 0.49326       & 0.3883        & 0.2701        & 0.7835      & 0.7881      & 0.8151      & 0.7956                                                                                                \\
\textbf{FightGliomas}          & 166                                           & 0.8275      & 0.7783      & 0.4583      & 0.3172        & 0.2312        & 0.51028       & 0.00239       & 0.0008        & 0.0007        & 0.8360      & 0.8488      & 0.6491      & 0.7779                                                                                                \\
\textbf{NIC-VICOROB}           & 166                                           & 0.3077      & 0.6883      & 0.6393      & 0.5380        & 0.0458        & 0.08012       & 0.00000       & 0.0000        & 0.0000        & 0.5899      & 0.8808      & 0.8531      & 0.7746                                                                                                \\
\textbf{LRDE\_2 \citep{LRDE2019}}               & 166                                           & 0.8851      & 0.8387      & 0.7725      & 0.5930        & 0.7017        & 0.26159       & 0.05312       & 0.0439        & 0.0196        & 0.7463      & 0.6977      & 0.8304      & 0.7581                                                                                                \\
\textbf{LRDE\_VGG \citep{LRDE2019}}             & 166                                           & 0.8810      & 0.7883      & 0.6303      & 0.4930        & 0.7313        & 0.83645       & 0.04460       & 0.0280        & 0.0185        & 0.7812      & 0.6764      & 0.5918      & 0.6831                                                                                                \\
\textbf{ANSIR}                 & 166                                           & 0.8727      & 0.8551      & 0.8349      & 0.0124        & 0.0765        & 0.11249       & 0.92500       & 0.9250        & 0.9250        & 0.6451      & 0.6179      & 0.5992      & 0.6207                                                                                                \\
\textbf{med\_vision \citep{medvision2019}}           & 166                                           & 0.8794      & 0.8512      & 0.8491      & 0.0203        & 0.0768        & 0.13209       & 0.92435       & 0.9253        & 0.9257        & 0.6449      & 0.6164      & 0.5971      & 0.6195                                                                                                \\
\textbf{TEAM\_ALPACA \citep{ALPACA2019}}          & 166                                           & 0.8768      & 0.8377      & 0.8116      & 0.0191        & 0.0707        & 0.10695       & 0.91639       & 0.9170        & 0.9228        & 0.6471      & 0.6167      & 0.5940      & 0.6192                                                                                                \\
\textbf{ODU\_vision\_lab}      & 166                                           & 0.8789      & 0.8517      & 0.8481      & 0.0212        & 0.0776        & 0.13283       & 0.92444       & 0.9253        & 0.9257        & 0.6444      & 0.6162      & 0.5965      & 0.6191                                                                                                \\
\textbf{DRAG \citep{DRAG2019}}                  & 161                                           & 0.8890      & 0.8518      & 0.8105      & 0.0726        & 0.1312        & 0.13792       & 0.92280       & 0.9241        & 0.9243        & 0.6312      & 0.5989      & 0.5828      & 0.6043                                                                                                \\ \hline
\end{tabular}%
}
\label{tab:QUBraTS2019}
\end{table}
\end{landscape}

\noindent

\end{document}